\documentclass[aps,pra,amsfonts,amssymb,twocolumn,floatfix]{revtex4}

\usepackage{graphicx}		
\usepackage{color}			
\usepackage{transparent}
\usepackage{dcolumn}			
\usepackage{booktabs}			
\usepackage{amsmath,array}
\usepackage{relsize}
\usepackage{enumerate}

\usepackage{listings}
\usepackage{epstopdf}

\begin{document}

\title{Classical crystal formation of dipoles in two dimensions}
\author{K.~K. Hansen}
\author{D.~V. Fedorov}
\author{A.~S. Jensen}
\author{N.~T. Zinner}
\affiliation{Institute of Physics and Astronomy, Aarhus University, DK-8000, Denmark}

\begin{abstract}
We consider a two-dimensional layer of dipolar particles in the regime of strong 
dipole moments. Here we can describe the system using classical methods and determine the crystal 
structure that minimizes the total energy. The dipoles are assumed to be aligned by an external 
field and we consider different orientations of the dipolar moments with respect to the two-dimensional 
plane of motion. We observe that when the orientation angle changes away from perpendicular and 
towards the plane, the crystal structure will change from a hexagonal form to one that has the 
dipoles sitting in equidistant rows, i.e. a striped configuration. In addition to calculating the 
crystal unit cell, we also consider the phonon spectrum and the speed of sound. As the 
orientation changes away from perpendicular the phonon spectrum develops local minima that are a 
result of the deformation to the crystal structure. 
\end{abstract}
\date{\today} 
\maketitle

\section{Introduction}
Quantum degenerate gases of dipolar particles is a forefront
research topic in cold atomic gases \cite{dipoleexp1,dipoleexp2,dipoleexp3,dipoleexp4,dipoleexp5,dipoleexp6,ni2010,ospelkaus2010,miranda2011,yan2013,zhu2014,aikawa2014,takekoshi2014}.
While losses can be a problem when dipolar forces are strong \cite{ni2010,ospelkaus2010,lushnikov2002,micheli2010}, it 
has been experimentally demonstrated that losses can be suppressed by
confinement in lower-dimensional geometries \cite{miranda2011}. 
This opens up the possibility of realizing interesting lower-dimensional 
phenomena driven by long-range interactions in both few-body 
\cite{fb1,fb2,fb3,fb4,fb5,fb6,fb7,fb8,fb9} and many-body \cite{mb1,mb2,mb3,mb4,mb5,mb6,mb7,mb8,mb9,mb10,mb11,mb12,mb13,mb14} physics. 
An overall goal is to realize a quantum simulator with long-range interactions
\cite{carr2009,tr1,tr2}. 

A recent drive in experiments has been to produce dipolar gases using 
heteronuclear molecules with large dipoles moments \cite{takekoshi2014,molony2014,shimasaki2014}. It is then 
expected that one can see strong dipolar effects already before reaching
degeneracy. When the dipolar energy scale exceeds the energy scale set by the
temperature of the system one may expect to see crystal formation in the 
classical sense. The regime where strong dipolar forces are the dominant 
feature of the system is the focus of the present paper, and we will 
thus ignore thermal effects. We will be considering
a geometry where the dipolar particles can move in two dimensions and 
where an external field is used to align the dipole moments at any fixed 
angle with respect to the plane of motion. We note that near the magic angle (to be
defined below) where two dipoles on a line will have vanishing dipolar
interaction energy, a normal mode in the crystal will approach zero
excitation energy. In this regime, we do except quantum and thermal 
fluctuations to play an important role, but leave this question for 
future studies.

In constrast to the famous Wigner crystals that have been predicted
for systems with Coulomb interactions at low density
\cite{wigner1934,bonsall1977}, the crystal phases should be found at
high density when the particles have dipolar interactions. Some
earlier works have considered dipoles oriented perpendicular to the
motional plane both classically \cite{kalia1981,bedanov1985,groh2001,lu2008,ramos2012}
and quantum mechanically \cite{dc1,dc2,dc3,dc4,dc5,dc6}. This leads to a
dipole-dipole force that only depends on the relative distance thus
strongly simplifying the problem. Another line of investigation has
been dipoles on a two-dimensional lattice
\cite{quin2009,carr2010,capo2010,gads2012}. Some of the latter works
have used different orientations of the dipoles with respect to the
lattice. In this case the dipole-dipole interactions no longer
cylindrically symmetry in the plane but can be highly anisotropic.
This may lead to striped systems as indicated in different response
function approaches \cite{sun2010,zinner2011,parish2012,mb14}.  It may
as well lead to new unexpected properties like the roton minimum in
helium \cite{helium}. We note that similar physical issue can be studied
using externally oriented magnetic colloids \cite{froltsov2003,froltsov2005}.

Naturally, these possibilities complicate both classical and quantum
mechanical approaches.  Classical properties of crystals are crucial
ingredients for subsequent quantization \cite{lindemann,chaikin}, and
investigations of quantum effects like melting point and heat capacity
\cite{halperin1978,nelson1979,chaikin,bruun2014}.  In the current work
we assume that the dipole moments are strong such that classical
methods are accurate and consider dipolar particles in a plane with no predefined
lattice, i.e. the particles can move continuously.  The question then
becomes whether, or perhaps more appropriately when, the crystal
structure starts to change significantly from the hexagonal crystal
structure that is found when the dipoles are aligned perpendicular to
the plane of motion.

The paper is organized as follows. In Sec.~\ref{methods} we describe
the theoretical model and the parametrization of the most general
crystal structure in the plane for arbitrary orientation of the
dipoles. Sec.~\ref{results} outlines the results for the crystal
structure itself.  In Sec.~\ref{phonon1} we elaborate on the
properties of the minimal energy crystal configuration. We calculate
the phonon spectrum by outlining a proper parametrization of the
reciprocal lattice before presenting the spectra along with the sound
velocity. Sec.~\ref{sum} contains a short summary and an outlook.  The
technical details of the calculations of the phonon spectrum are
presented in Appendix~\ref{appa}.

\begin{figure}
\centering
\includegraphics[scale=0.7]{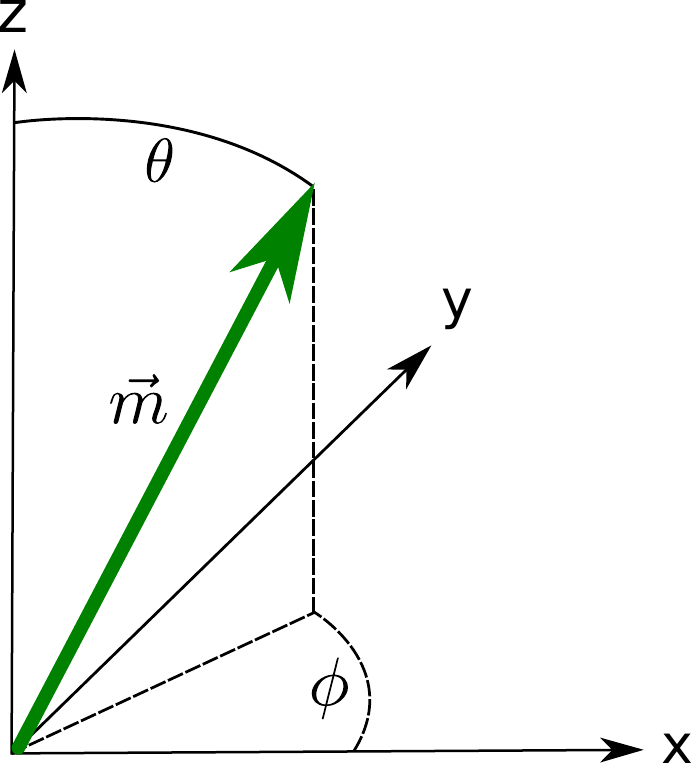}
\caption{The angular coordinates, $\theta$ and $\phi$, describing the
  direction of the dipole moment, $\vec{m}$.  }
\label{fig:dipol}
\end{figure}

\section{Crystal parametrizations}\label{methods}
In order to determine the crystal structures for different aligned
dipole moments of the particles, we need to develop a convenient
parametrization of the geometry.  We imagine a system with a number of
particles distributed on a section of a plane in lattice-like
structures.  Instead of the number of particles in the allowed space
it is much better to use the density of particles (an inverse area in
2D) as a decisive parameter.  The anticipated regular structure can
then easily be broken down into unit cells repeating each other across
the plane.  The choice of parametrization is intrinsically connected
to the interactions between the particles.  We therefore first examine
the details of the dipole-dipole potential and subsequently in the next
subsection we describe the parametrization of the unit cell.

\subsection{Dipole-dipole interaction}

All dipoles are assumed to be identical with a given dipole moment of
size $D$.  They all have the same orientation induced by the
application of an external magnetic or electric field.  The direction
of the dipole moments is given by the angles $(\theta$, $\phi$) as
illustrated in Fig.~\ref{fig:dipol}. The dipole moment itself,
$\vec{m}$ is then described by
\begin{align}
\vec{m} = D 
\begin{pmatrix}
\sin (\theta) \cos(\phi) \\ \sin(\theta) \sin(\phi) \\ \cos(\theta)
\end{pmatrix}.
\end{align}
The interaction energy between two of these dipoles, $\vec{m}_1$ and
$\vec{m}_2$, is
\begin{equation}
I = C \left( \frac{\vec{m}_1\cdot\vec{m}_2}{r^3}-\frac{3 \left( \vec{m}_1\cdot\vec{r}\right)\left( \vec{m}_2\cdot\vec{r}\right)}{r^5} \right),
\label{eq:interaction}
\end{equation}
where $\vec{r}$ is the vector between the two dipoles, $r=|\vec{r}|$,
and $C=\tfrac{\mu_0}{4\pi}$ and $C=\tfrac{1}{4\pi\epsilon_0}$ for
magnetic and electric dipoles, respectively.   Here $\mu_0$ and
$\epsilon_0$ are the vacuum permeability and vacuum permittivity,
respectively. Below we will assume magnetic dipoles. Results for 
electric dipoles can be obtained by simply making the substitution
$\mu_0\to 1/\epsilon_0$.

\begin{figure}
\centering
\includegraphics[scale=0.25]{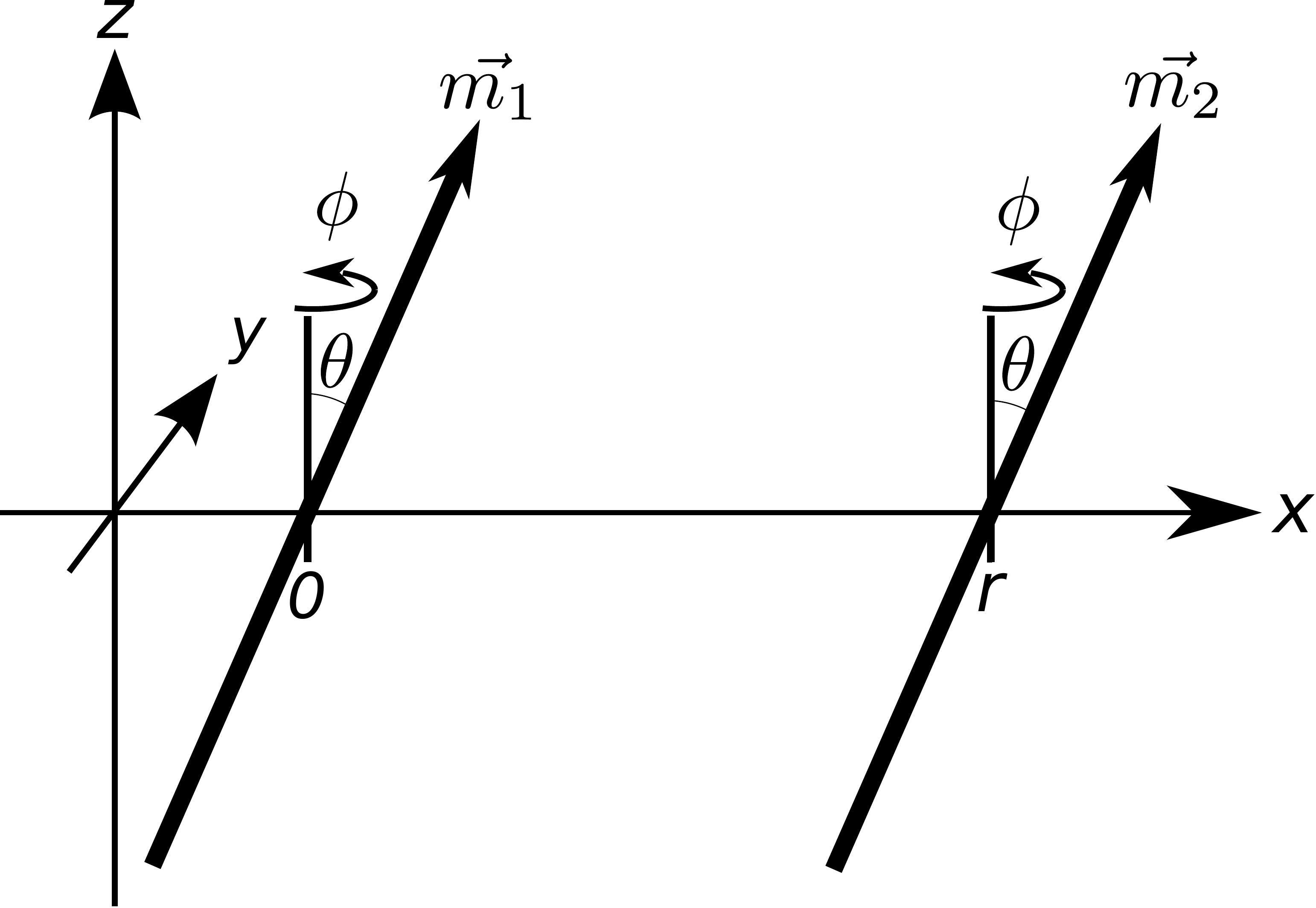}
\caption{Illustration of two identical aligned dipoles on the $x$-axis
  separated a distance $r$ from each other.  Both directions are
  described by the same $(\theta,\phi)$ as in Fig.~\ref{fig:dipol}.  }
\label{fig:interaktion}
\end{figure}

We now assume that the two dipoles are placed on the $x$-axis at a
distance $r$ and described by the angles $\theta$ and $\phi$. For
identical dipole moments $(\vec{m}_1=\vec{m}_2)=\vec{m}$ the potential
energy in Fig.~\ref{fig:interaktion} is reduced to
\begin{equation} \label{eq: todipoler}
I = C \left( \frac{D^2}{r^3}-\frac{3 \left( \vec{m}\cdot\vec{r}\right)^2}{r^5} \right)=
\frac{C D^2}{x^3}\left( 1-3\sin^2\theta\cos^2\phi \right),
\end{equation}
since $\vec{r}$ is taken along the $x$-axis.  We thus see that the
interaction energy is positive as long as
\begin{equation}
1-3\sin^2\theta \cos^2 \phi >  0.
\end{equation}
We therefore conclude that the interaction is repulsive for all $\phi$
when $\theta < \theta_c \equiv \sin ^{-1} \left(1 / \sqrt{3} \right)=
0.61548$.  Thus, collapse due to attractive inverse cubic
interactions can not occur for $\theta < 0.61548$.  We note that the
$\sin^2\theta$ dependence implies a symmetry around $\theta= \pi /2$
corresponding to reflection in the $xy$-plane where all dipoles are
assumed to be located.

\subsection{Unit cell}

When comparing different crystal structures we must consider the
conditions we impose on the system.  We imagine that the crystal
forms from a gas phase that condenses by cooling.  Consequently the
crystal structures most likely prefer to have a constant density of
dipoles in the plane, i.e. we do not have clumping or holes in the
crystal.  We thus assume an ideal uniform crystal structure.  It is of
course very interesting to study the effect of crystal defects in both
a classical and quantum mechanical setting but this is beyond the
scope of the present study.

We therefore aim at determining the preferred regular crystal
structure for a given alignment of the many dipole moments distributed
to produce a given average density in the $xy$-plane.  We then need
to find the configuration that minimizes the total interaction energy
for such specification of the system.  The total interaction energy,
$U$, is
\begin{equation} \label{eq:energy}
U = \frac{1}{2} \sum_{\substack{
\vec{R}\vec{R'} \\ \vec{R} \neq \vec{R'}}}
I (\vec{R}-\vec{R'}) \; ,
\end{equation}
where $\vec{R}$ and $\vec{R'}$ are position vectors of each pair of
dipoles interacting through the potential, $I(\vec{r})$, in
Eq.~\eqref{eq:interaction}.

We need a sensible unit cell that can be used to parametrize the
crystal structure for given external alignment.  An important limit to
consider is the one of perpendicular dipoles, i.e.  when all the
dipole moments are aligned perpendicular ($\theta=0$) to the plane
containing all the particles.  In this case the dipole-dipole
interaction is purely repulsive with a $r^{-3}$ behavior that only
depends on the relative distance.  The system will try to
maximize the distance between each pair of dipoles and not allow any
clusterization or irregular correlations (at least in the ideal
crystal we investigate).  Finding the minimum energy configuration for
this system is therefore the same problem as the packing of spheres.
This problem is known to have a hexagonal solution in 2D \cite{faststof2}and we therefore must be able to capture this geometry with our unit
cell parametrization.

\begin{figure}
\centering
\includegraphics[scale=0.25]{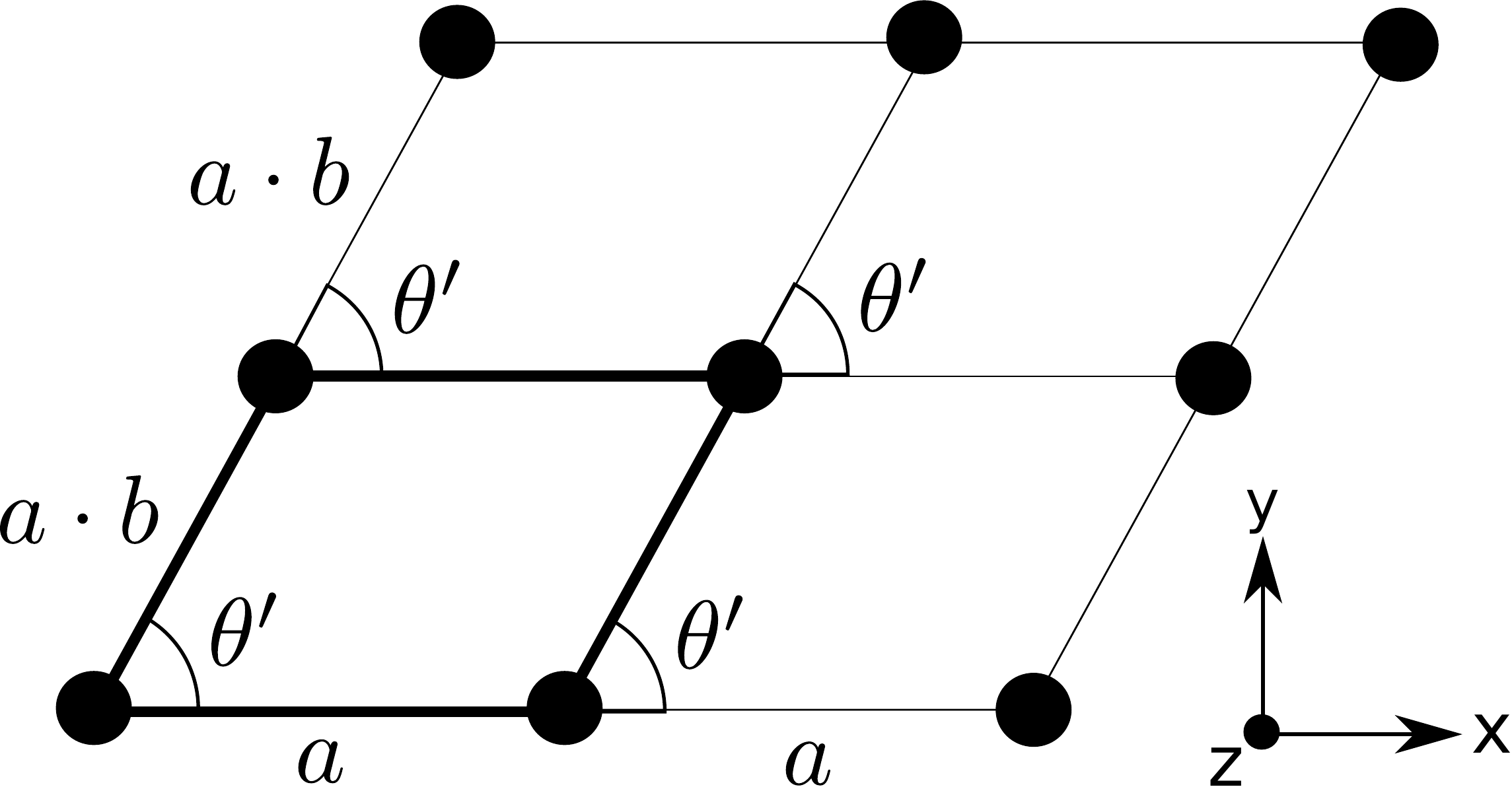}
\caption{Illustration of the parametrization used for the unit cell in
  the crystal $xy$-plane where the $z$-axis is perpendicular to the
  plane with the particles. The unit cell is shown in the bottom 
	left corner with bold lines. 
	The parametrization contains two
  variables, $\theta'$ and $b$. $\theta'$ determines the angle in the lattice, $a$
  and $ab$ are the distance between nearest neighbors in the $x$- and
  the $\theta'$-directions, respectively.  With this choice we can
  parametrize all uniform crystal structures.}
\label{fig:opstilling}
\end{figure}

To allow sufficient flexibility we choose the unit cell shown in
Fig.~\ref{fig:opstilling}.  The two variables $\theta$' and $b$ allow
us to describe all uniform crystal structures with $a$ as the crystal
lattice length.  Symmetry allows a restriction of $\theta$' to the
interval $[0,\pi/2[$ and $b$ to be less than unity, i.e. $]0,1]$.  The
value of $a$ can be determined as function of $\theta$' and $b$ to
maintain a given average density of dipoles in the resulting crystal.
This can be imposed by insisting that the unit cell area remains
constant as we change the parameters $\theta'$ and $b$. The density,
$\sigma$, is inversely proportional to the area, $a^2 b\sin\theta'$,
of the unit cell.  The constraint on $a$ is then
\begin{equation}
a = \sqrt{\frac{1}{\sigma b\sin\theta '}}.
\end{equation}
The two unit cell basis vectors, $a_i$, connecting nearest neighbors
in the two principal directions are
\begin{equation}
\vec{a}_1 =
\begin{pmatrix}
a \\ 0
\end{pmatrix}
, \vec{a}_2 = 
\begin{pmatrix}
ab\cos\theta' \\ ab\sin\theta'
\end{pmatrix}
\label{eq: gittervektorer}.
\end{equation}

We may now exploit the symmetries of the unit cell.  As mentioned
before there is a symmetry around $\theta = \pi/2$.  There is also a
symmetry resulting in $U\left(\phi\right)=U\left(\phi+\pi\right)$
where the energy is invariant under translation of $\phi$ by $\pi$.  This can
easily be seen by the invariance of rotating the unit cell 180 degrees
around the $z$-axis, e.g. rotating around the central point in
Fig.~\ref{fig:opstilling}.  It can also be seen by looking at the
expression for the dipole interaction in Eq.~\eqref{eq: todipoler}.
The dependence on $\phi$ goes as $\cos^2\phi$ and this is invariant
under rotations by $\pi$.  We can therefore reduce the parameter space
and work with $\theta \in [0,\pi/2]$ and $\phi \in [0,\pi]$.

It is important to note that the external alignment of the dipoles
fixes $\theta$ as the angle between the $z$-axis and the direction of
the external electric or magnetic field.  The other angle, $\phi$,
describes the angle between the arbitrarily chosen $x$-axis of the
unit cell and the projection on the crystal plane of the aligned
dipole moments.  The calculational procedure will therefore be to fix
$\theta$ and $\phi$, and then minimize the energy with respect to
$\theta'$ and $b$.  This will for each $\theta$ produce an energy as a
function of $\phi$ and the minimum will then define the preferred
crystal structure for this given $\theta$.

In general a complicated topology may invalidate this procedure of
fixing one parameter, minimizing with respect to two other parameters,
and afterwards find minimum of the resulting function of the first
parameter.  All three variables should perhaps be varied
simultaneously to find true global or local minima.  To confirm that
our procedure provided correct minima we tested by numerical variation
of the parameters around the minima.

\section{Crystal structure}\label{results}
The structure of uniform crystals with constant planar density of
dipoles has been determined as function of polarization angles.  We
shall here present the resulting configurations for a number of angles
that illustrate the general behavior.  We will start with $\theta=0$
(independent of $\phi$).  Subsequently we increase $\theta$ up to the critical
value, $\theta_c$, for collapse due to unhindered small-distance
attraction.  We shall investigate how the crystal evolves for
different external alignments of the dipoles, that is for the interval
$0<\theta<\theta_c$ where $\phi$, $\theta'$ and $b$ assume correlated
values minimizing the total energy.  We emphasize again that $\theta'$
and $b$ describe the crystal structure in Fig.~\ref{fig:opstilling}
whereas $\phi$ describe the orientation of the crystal in the plane
with respect to the external polarization direction.

\begin{figure}
\centering
\includegraphics[scale=0.5]{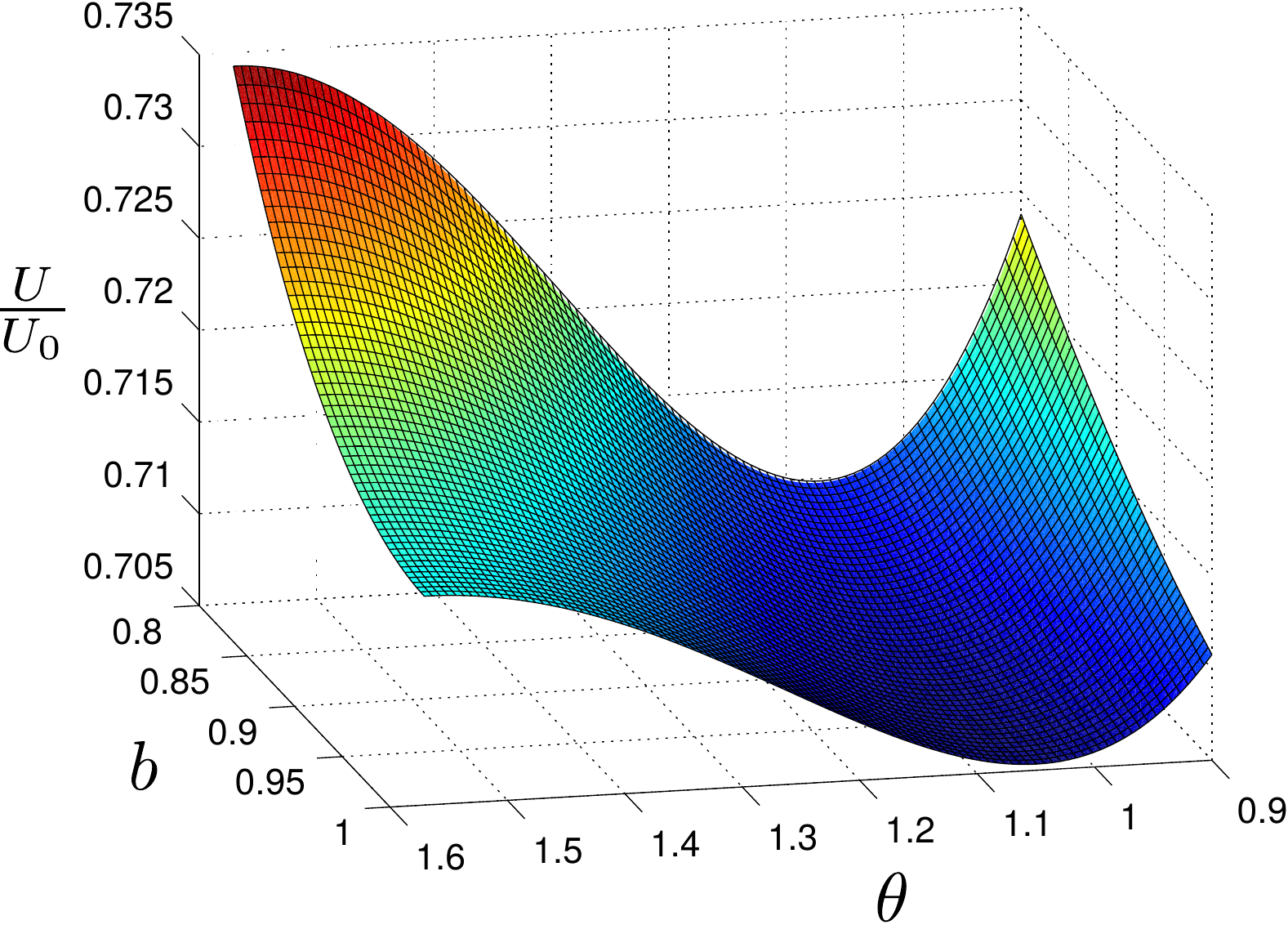}
\caption{The total energy for $\theta = 0$ (independent of $\phi$) as
  a function of $b$ and $\theta '$ is here shown in units of $U_0 =
  \mu_0 D^2 \sigma ^{3/2}$. The constant two-dimensional density,
  $\sigma$, is an inverse area and consequently $\sigma ^{3/2}$ is an
  inverse cubic length.  Only a section of the allowed values of $b$
  and $\theta '$ is shown as it is easily seen that the energy
  increases rapidly beyond what is shown. The lowest values are dark
  blue and increasing energies follow the colors of the rainbow. }
\label{fig: overflade}
\end{figure}

\begin{figure}
\centering
\includegraphics[scale=0.25]{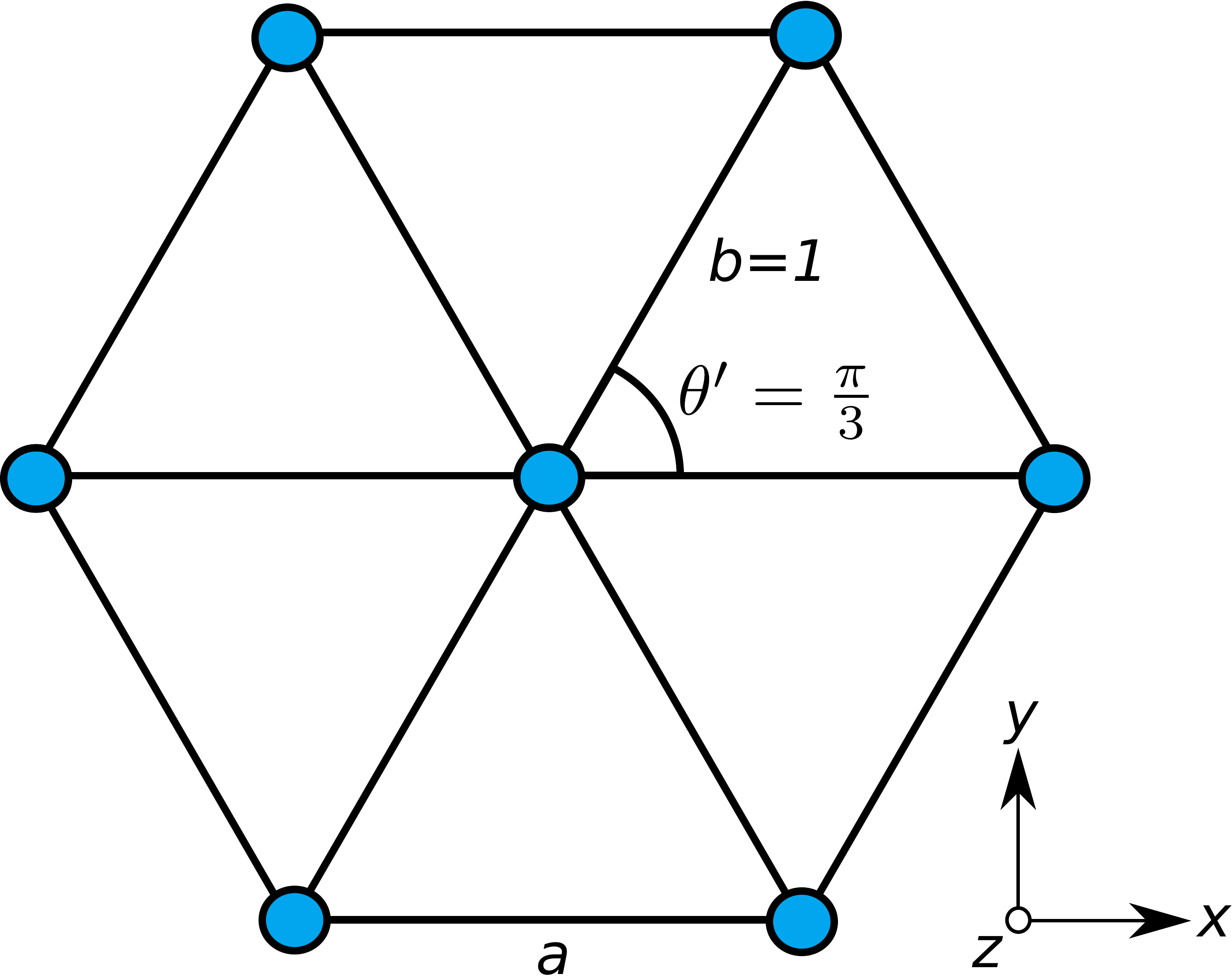}
\caption{The hexagonal lattice structure for $\theta=0$ with the
  specified crystal parameters, $\theta'$ and $b$, as defined in
  Fig.~\ref{fig:opstilling}. }
\label{fig: trivistruk}
\end{figure}

\subsection{Perpendicular dipoles: $\theta=0$}
First we consider the structures arising for $\theta=0$ where the
dependence on $\phi$ drops out.  It is indeed the simplest case yet it
is a relevant benchmark to demonstrate the validity of our method.
The total interaction energy from Eq.~\eqref{eq:energy} as a function
of $\theta'$ and $b$ is shown in Fig.~\ref{fig: overflade}.  Energies
are plotted in units of $U_0 = \mu_0 D^2 \sigma ^{3/2}$ where the
density of dipoles, $\sigma$, is kept constant in all calculations.

As expected, we find that the minimum in energy occurs when $\theta' =
\pi/3$ and $b=1$. The unit cell of the corresponding hexagonal lattice
structure of lowest energy is shown in Fig.~\ref{fig: trivistruk}.
Note that for $\theta=0$ there is a triangular symmetry which is the
result of the rotational symmetry in the plane for $\theta=0$ which is
broken down to the point-group symmetry of the hexagonal lattice when
the crystal forms.

A rather flat valley extends in Fig.~\ref{fig: overflade} from the
minimum in the direction of smaller $b$ for fixed $\theta' = \pi/3$.
The walls increase rather steeply on both sides of this valley.  The
crystal is therefore relatively soft towards moving the upper (and
lower) dipoles in Fig.~\ref{fig: trivistruk} further away from (or
closer to) each other and the central line of dipoles along the
direction of $\theta'=\pi/3$.  We emphasize that precisely the same
unit cell is obtained for $\theta'=2\pi/3$.  Consequently a similar
minimum would show up by extending Fig.~\ref{fig: overflade} to larger
values of $\theta'$. The barrier between the two minima is already
indicated and almost reached.  No other direction parametrized by
$\theta'$ provide the same structure for the same density.

Small changes of $\theta$ away from zero would break the hexagonal
lattice symmetry and introduce a preferred direction connected to the
direction of the external field that aligns the dipoles.  The soft
directions along $\theta'=\pi/3$ and $\theta'=2\pi/3$ then must be
either followed or overpowered. The hexagonal structure must be
distorted or completely changed.

\begin{figure}
\centering
\includegraphics[scale=0.5]{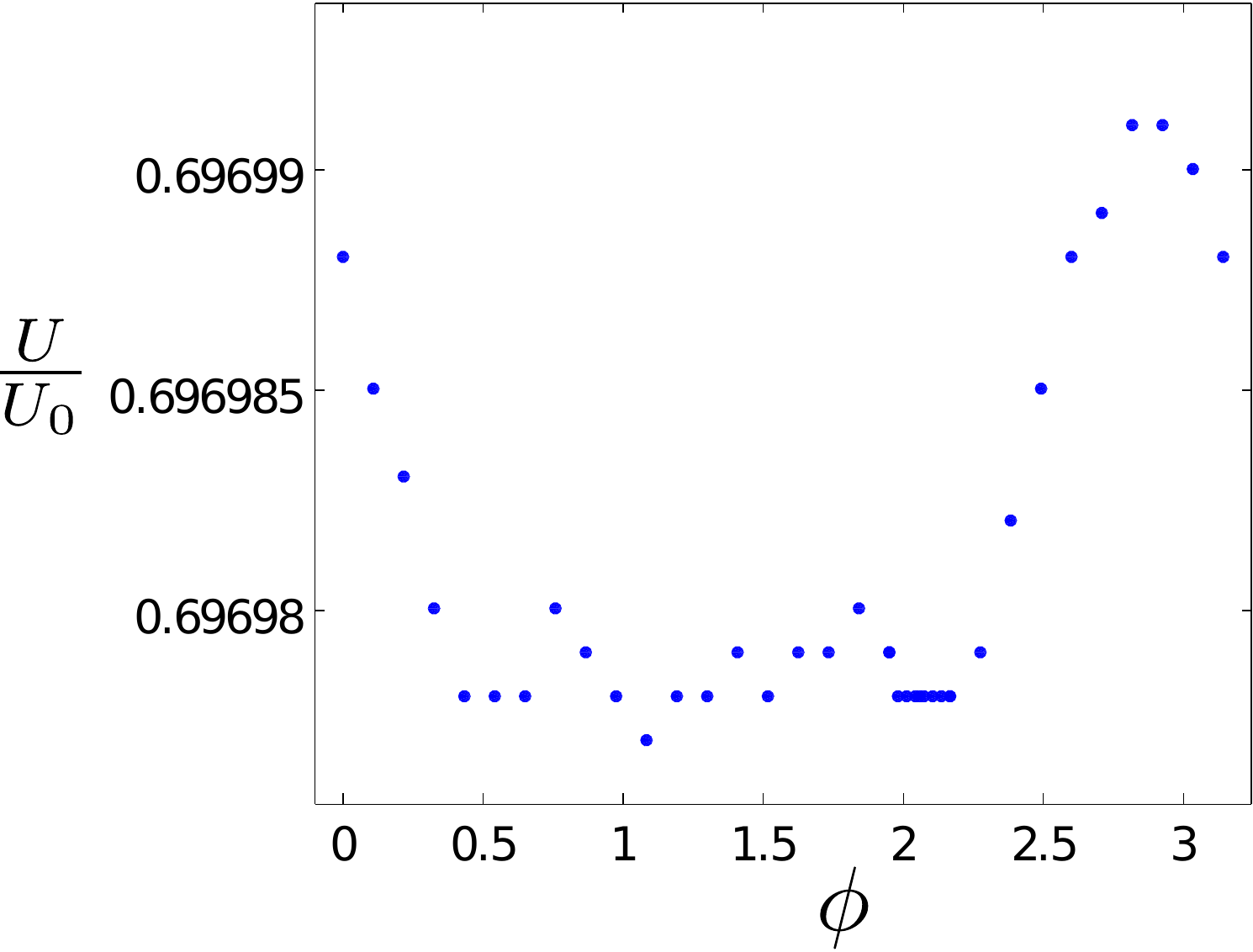}\\
\includegraphics[scale=0.5]{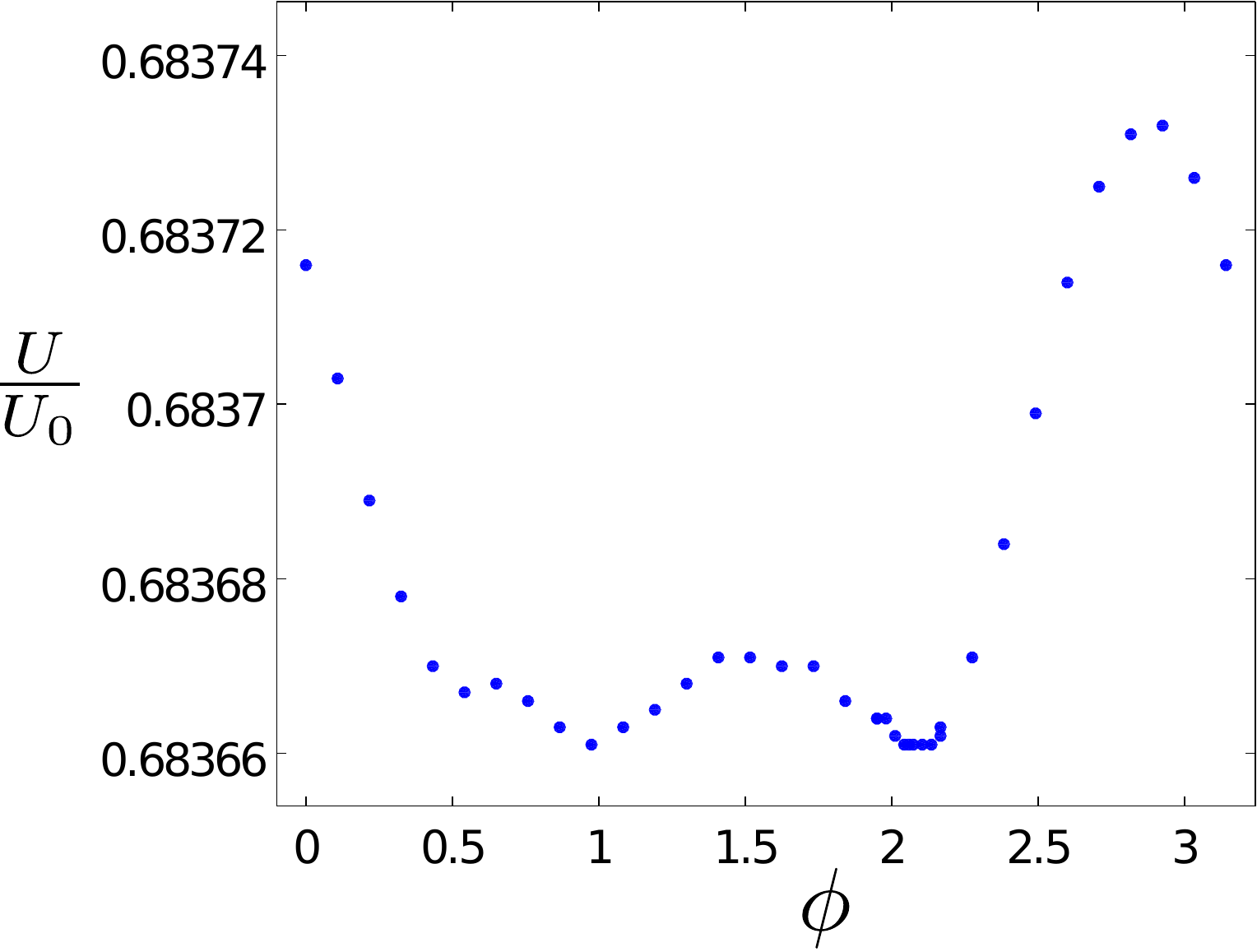}\\
\includegraphics[scale=0.5]{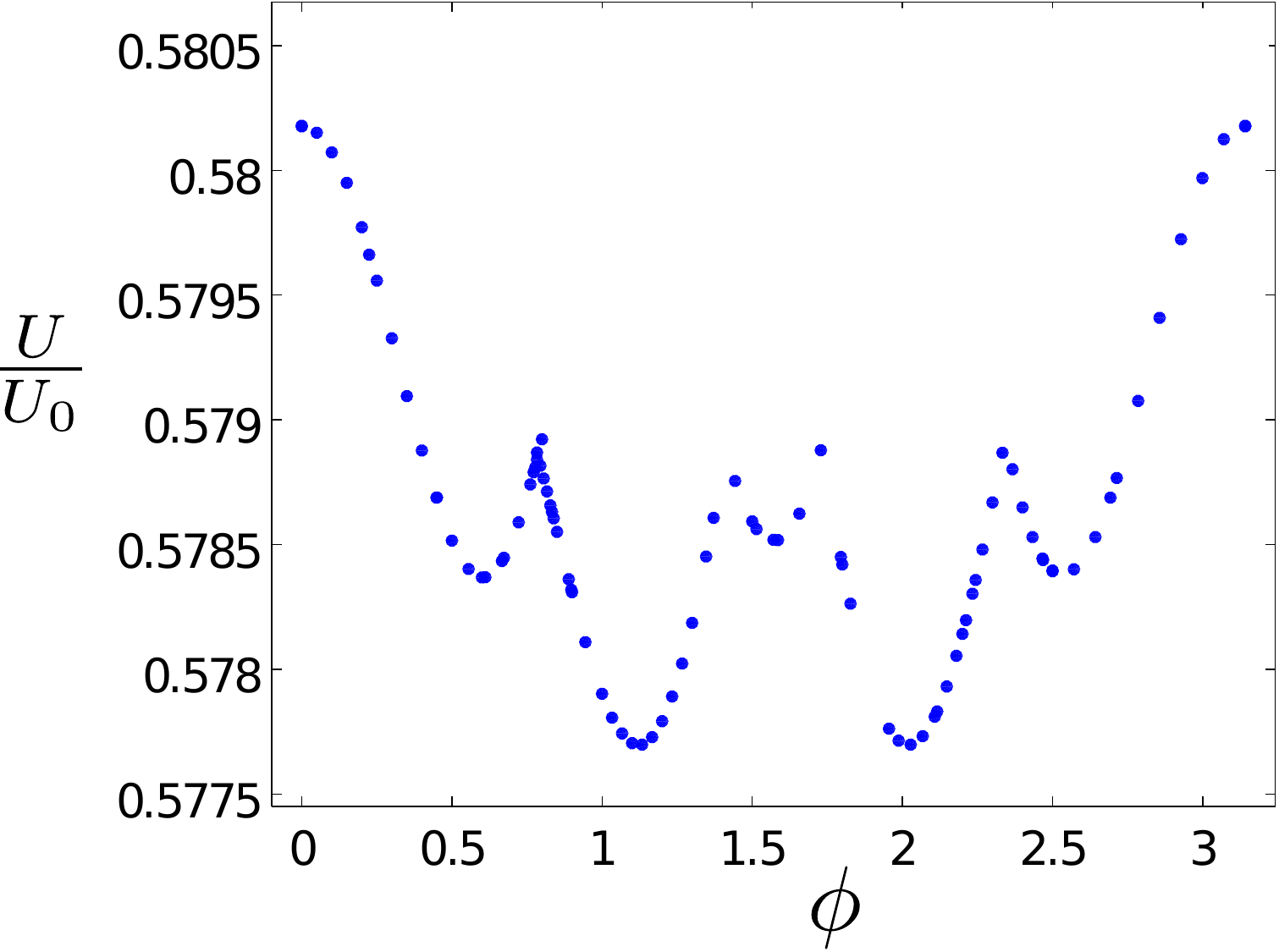}
\caption{The interaction energy in Eq.~\eqref{eq:energy} in units of
  $U_0$ defined in Fig.~\ref{fig: overflade} for
  $\theta=0.1,0.15,0.35$ (upper, middle, lower panels) as a function of
  $\phi$ after minimization with respect to $\theta'$ and $b$. }
\label{fig: cirkel0,1}
\end{figure}

\subsection{Tilted dipoles: $0.1 \leq \theta \leq 0.6$}
We now move the polarization direction, $\theta$, away from being
perpendicular to the crystal plane.  The crystal has to adjust by
finding the minimum energy with respect to $b$, $\theta'$, and $\phi$.
For convenience we minimize with respect to $b$ and $\theta'$ for each
value of $\phi$.  The resulting functions are shown in Fig.~\ref{fig:
  cirkel0,1} for three different values of $\theta$.

The energy for the smallest polarization angle, $\theta=0.1$, is
exhibited in the upper panel.  The variation with $\phi$ is
exceedingly small with a decrease from the $\theta=0$ result of
$0.7072$ by around $1\%$.

The case of small angle polarization is nevertheless the first step
away from the highest symmetry and towards collapse of the crystal
structure.  The case of $\theta=0.1$  in Fig.~\ref{fig: cirkel0,1} already
exhibits all the tendencies to develop pronounced minima in the
potential energy surface.  We observe the global minimum at $\phi =
1.083$, which corresponds to $\theta' = 1.056$ and $b=0.99$.
The structure for this minimum is then visualized as a polarization
direction, $\phi$, tilted in the direction of $\theta' = \pi/3$.

We also note additional (local) minima around $\phi = 2.058$ and
$\phi = 0.542$.  The first of these corresponds to a structure with
$\phi= \theta' = 2 \pi/3$, that is the same geometry as the global
minimum but for a different unit cell.  The second minimum corresponds
to $\phi \approx \pi/6$, $\theta' \approx \pi/3$, and $b \approx
0.99$, that is an apparently metastable structure.  We emphasize that
all minima are tested to be real minima by direct computation of the
energy surface in a cubic grid around the minimum points.  In any case
the energy differences are very small and sometimes at the limit of
being significant.

Still it is of interest to understand the emerging structures.  The
minimum energy configuration for $\phi =0.542$ is the same as for
the two deeper minima but now the polarization direction is half way
between those of the preferred global minimum, $\phi \approx \theta' =
\pi/3$, and the extreme linear structure, $\phi \approx \theta' = 0$.
The symmetry is very high as the polarization points directly to
towards the opposite, furthest away, dipole in the unit cell.

Going away from the flat region for intermediate $\phi$-values, both
towards smaller and larger values of $\phi$ we find a steeply
increasing energy.  The curve beyond $\phi= \pi$ continues precisely
as for $\phi=0$.  The maximum is relatively high and associated with a
structure corresponding to the hexagonal structure which is
energetically unfavored.

Tilting the dipole to $\theta=0.15$ leads to larger variation of the
potential energy as shown in the middle part of Fig.~\ref{fig:
  cirkel0,1}.  The deepest minimum associated with the optimal crystal
structure still appears for about the same structure values, $\phi =
2.058$, $\theta' = 1.025$ and $b = 1$, as for $\theta
= 0.1$.  The minimum with the same crystal structure for about $\phi =
0.975$, $\theta' = 1.067$ and $b = 0.98$ is fortunately also found with
the same energy.  The overall features in the energy as function
$\phi$ remain the same as for $\theta=0.1$ but now determined with
better relative accuracy.  The local minimum at about $\phi \approx
0.542$ has almost disappeared.

Proceeding to the larger polarization angle of $\theta=0.35$ we find
the energy shown in the lower part of Fig.~\ref{fig: cirkel0,1}.  The
two optimal minima are now rather pronounced at the same $\phi$-values
of $\phi = 2.028$ and $1.133$.  The two local minima at about
$\phi = 0.600$ and $2.500$ are now clearly seen in the figure.
They correspond to crystal structures described by $\theta' \approx 1$
and $b \approx 0.35$.  These local minima are sensitive to the influence
from dipoles at very large distances which tend to increase the
minimum values and perhaps eventually wipe them out completely.

\begin{figure}
\centering
\includegraphics[scale=0.5]{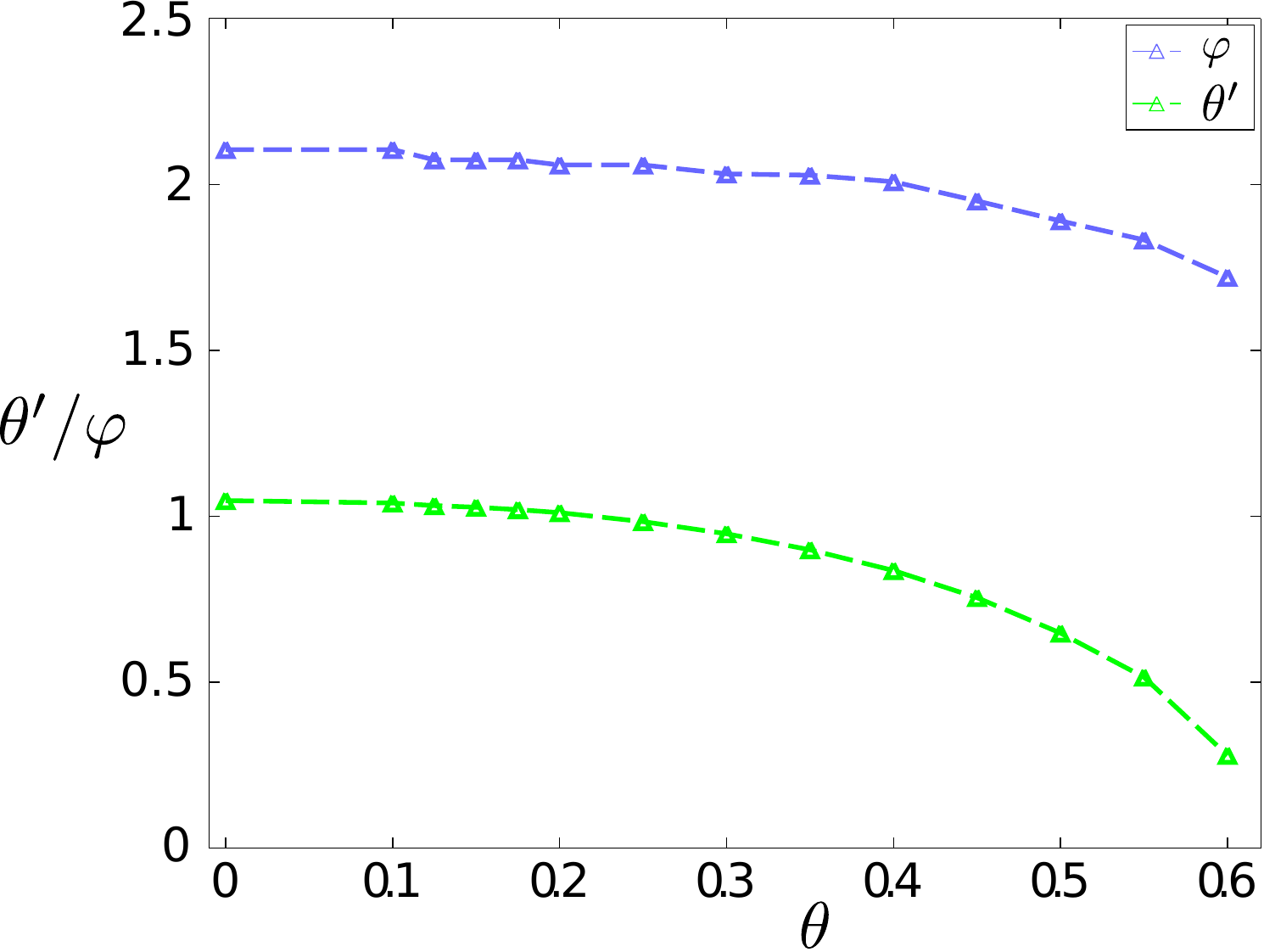}
\caption{$\theta'$ and $\phi$ is here shown for the global minima as a function of $\theta$.}
\label{fig: cirkel2}
\end{figure}

The minimum energy configurations for the different $\theta$-values
all correspond to $b \approx 1$.  The variations of $\theta'$ and
$\phi$ as functions of $\theta$ are shown in Fig. \ref{fig: cirkel2}.
The similarity is striking as the two equivalent minima correspond to
$\phi \approx \theta'$ and $ \phi \approx 2 \theta'$.  As already
mentioned this is not surprising as the same structure arises and the
polarization direction points along the highest symmetry axis of the
crystal structure.

\begin{figure}
\centering
\includegraphics[scale=0.3]{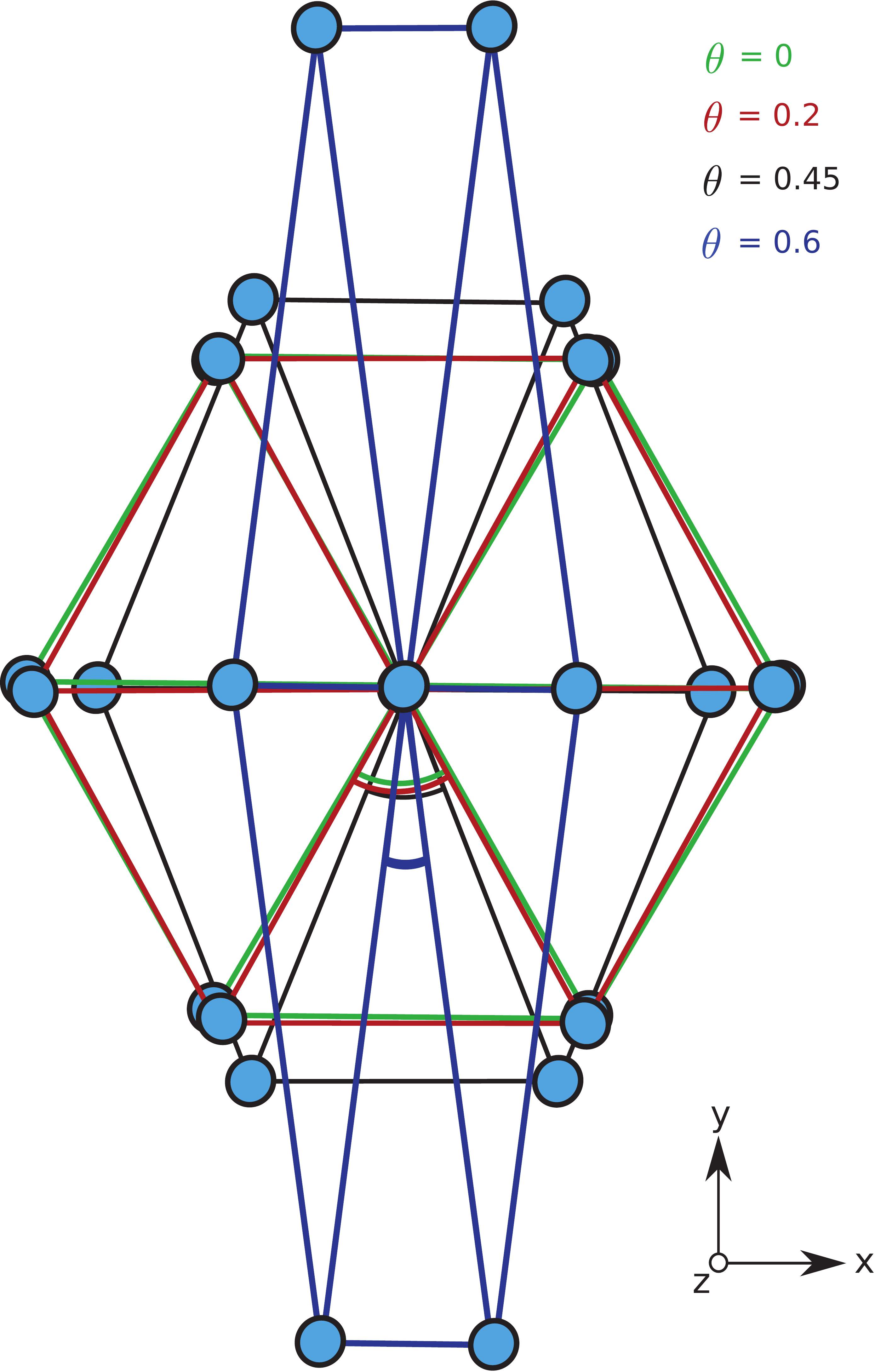}
\caption{The structure for $\theta = 0, 0.2, 0.45$ and $0.6$. The
  color fits with the structures shown. }
\label{fig: struktur-0-0,6}
\end{figure}

The two identical optimal structure changes systematically with
$\theta$ as illustrated in Fig. \ref{fig: struktur-0-0,6}.  It is
seen clearly that the structure evolves continuously from the
hexagonal structure in Fig. \ref{fig: trivistruk} and towards lines of
well separated dipoles.  The ultimate linear configurations place the
dipoles at pairwise minimum energy positions with respect to each
other.

The change in structure seems to be accelerating as $\theta$ becomes
larger and approaching the structure where each row of dipoles would
collapse on itself after crossing the threshold into the range of
inverse cubic attraction at zero distance. This configuration is very
near the point $\theta_c = 0.615$ where attraction between two dipoles
would occur, and it would seem logical that this is the point where
the structure would collapse.  The last configuration we investigated
was for $\theta = 0.6$ where the angle $\theta'$ is approaching the 
dive down towards zero as seen in in Fig. \ref{fig: cirkel2}.

The overall change in the system makes sense intuitively. As the
dipole-interaction weakens as $\theta$ grows it makes sense that the
dipoles would try to place themselves behind each other at the minimum
energy positions with the nearest neighbours. The more it weakens the
more favourable this positioning becomes and the structure will change
accordingly.

\section{Phonon spectra}\label{phonon1}
When the structure has been determined for a configuration of the
dipoles it will be of interest to determine the phonon spectrum.  This
requires calculations of the frequencies of the normal modes of the
lattice vibrations around stable minimum configurations.  We shall
first indicate the theoretical derivation and provide expressions and
calculational procedure for spectra and corresponding speed of sound.
The details can be found in Appendix~\ref{appa}.  
Second we exhibit the symmetries of the reciprocal
lattice as function of $\theta'$ and $b$. Finally we discuss the
calculated frequencies, for different polarization angles, $\theta$,
as functions of wave number in the two normal mode directions.

\subsection{Formulation}

The interaction energy is calculated by use of
Eqs.~\eqref{eq:interaction} and \eqref{eq:energy} for a given
equilibrium structure where the dipoles are located at a series of
lattice points, $\vec{R}_0$.  We move the dipoles a small distance
away from their respective equilibrium positions,
$\vec{u}(\vec{R}_0)$, such that $\vec{R} = \vec{R}_0 +
\vec{u}(\vec{R}_0)$. Then change of the pairwise distance becomes
$\vec{\delta u} = \vec{u}(\vec{R}_0) -\vec{u}(\vec{R}'_0)$ which
assumed small allows expansion of the energy in Eq.~\eqref{eq:energy}
to second order in $\vec{\delta u}$.  The zeroth order term will not
contribute to the dynamics. The first order term will vanish as we
expand around minimum positions.

We now assume equilibrium at $\vec{R}$ and omit from now on the index,
$''0''$.  The gradient of the second order term provides the force on
the amplitude, $\vec{u}$, and the equation of motion becomes
\begin{equation} \label{eq:motion}
M \ddot{\vec{u}} (\vec{R}) = -\vec{\nabla}_{\vec{u}(\vec{R})}  
I({\vec{u}(\vec{R})}) =
- \sum_{\vec{R'}} \underline{\underline{D}} (\vec{R}- \vec{R'}) \vec{u}(\vec{R'}),
\end{equation}
where $M$ is the mass of one dipole, and the elements of the
$D$-matrix are defined by
\begin{equation}
D_{\mu \nu} (\vec{R}-\vec{R'}) = \left. \frac{\partial^2 I}{\partial u_{\mu} (\vec{R}) \partial u_\nu (\vec{R'})}\right|_{u \equiv 0}.
\end{equation}
We search for periodic solutions, $\vec{u}(\vec{R})$, in a
direction described by the unit vector, $\vec{\epsilon}$, by inserting
the ansatz,
\begin{equation}
\vec{u}(\vec{R}) = \vec{\epsilon} e^{i(\vec{k} \cdot \vec{R} -\omega t)},
\end{equation}
into Eq.~\eqref{eq:motion} for a constant given wave vector,
$\vec{k}$.  The resulting two-dimensional eigenvalue equation gives us
two solutions corresponding to vibrational frequencies,
$\omega(\vec{k})$, and related to the two normal mode directions,
$\vec{\epsilon}$, parallel and perpendicular to the wave vector,
$\vec{k}$.

\begin{figure}[h]
\begin{minipage}{0.48\textwidth}
\centering
\includegraphics[scale=0.5]{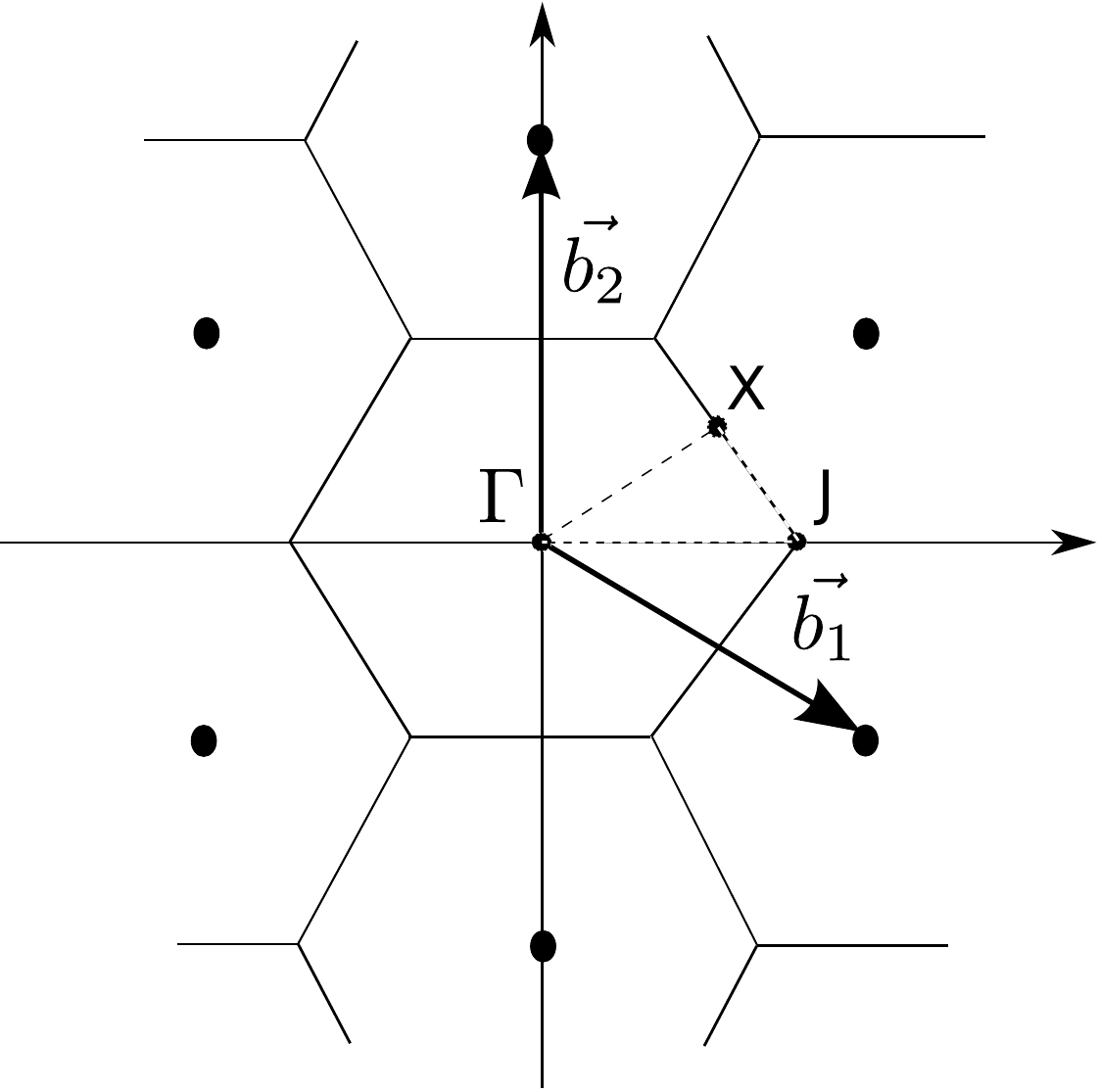} \\
\vspace*{0.5cm}
\includegraphics[scale=0.5]{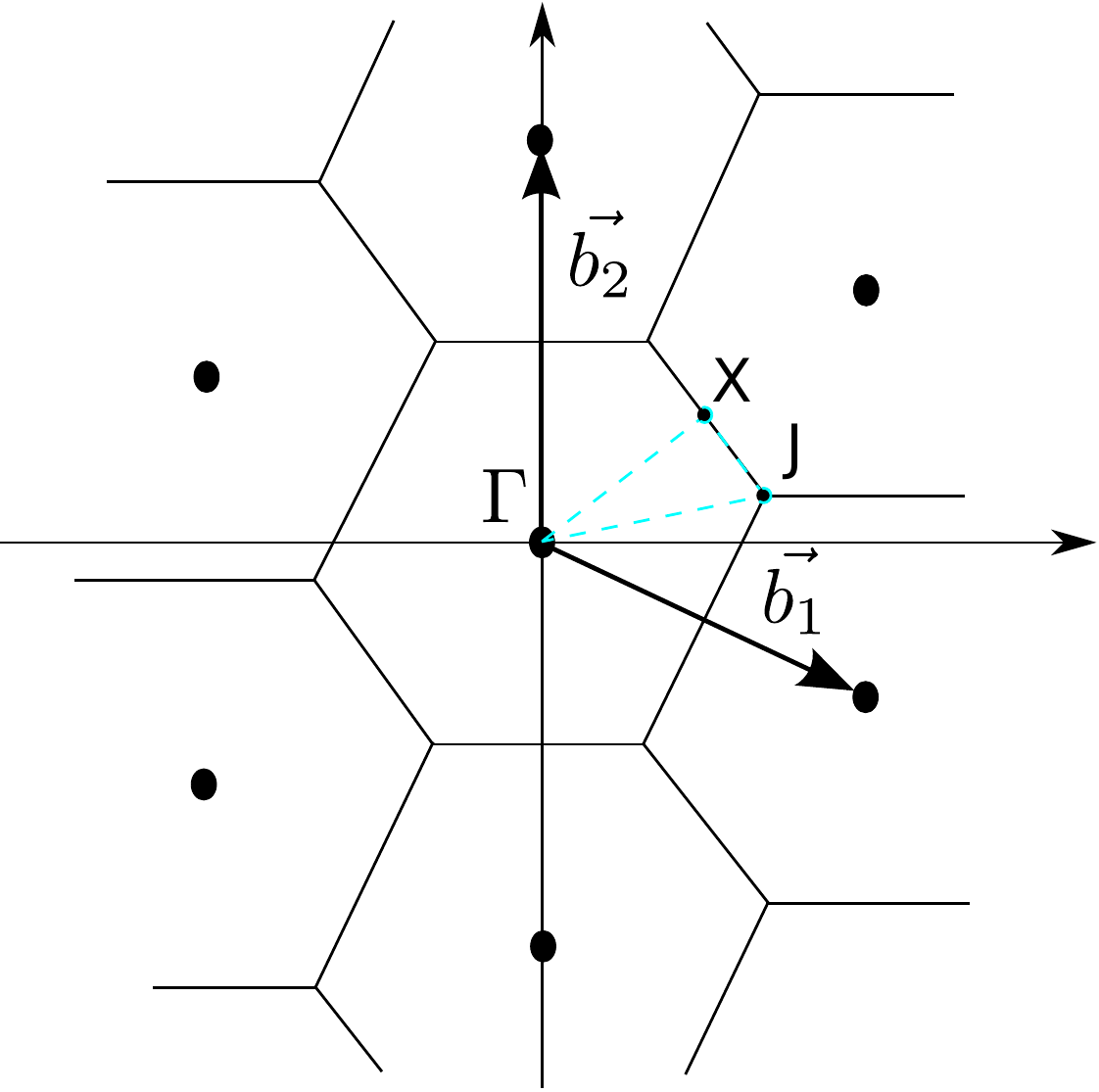}
\caption{The reciprocal lattices for $\theta = 0$ corresponding to
  $\theta ' = \pi/3$ and $b = 1$ (upper panel), and $\theta ' > \pi/3$
  and/or $b > 1$ (lower panel.  The lines represent the first
  Brillouin zone for each particle. Symmetry points at $\Gamma$, X and
  J are observed. These are respectively the point between two dipoles
  and the corner between three Brillouin zones.}
\label{fig: reci trivi}
\end{minipage}
\end{figure}

From the eigen frequencies determined for the central symmetry point,
$\Gamma$ in Fig. \ref{fig: reci trivi}, we calculate the corresponding
sound velocities,
\begin{align}
v = \left. \frac{\partial \omega (\vec{k})}{\partial \vec{k}} \right|_{\vec{k}=0},
\label{eq:lyd}
\end{align}
for the two normal modes, transverse and longitudinal waves.

\subsection{The Reciprocal Lattice}
To analyze the phonon spectrum we need to know the points of interest
in the reciprocal lattice. We therefore calculate the reciprocal
lattice as a function of $\theta '$ and $b$.  We define $\vec{b}_1$ an
$\vec{b}_2$ as the reciprocal lattice vectors. From the definition of
reciprocal lattice vectors we have
\begin{align}
\vec{a}_i\cdot \vec{b}_j = \delta_{ij} 2 \pi,
\end{align}
where $\vec{a}_i$ are the lattice vectors in Eq.\eqref{eq: gittervektorer}.
This leads to:
\begin{align}
\vec{b}_1 = \begin{pmatrix} \frac{2 \pi}{a} \\[1ex] \frac{-2 \pi}{a \tan \theta '} \end{pmatrix},
\hspace{0.3cm}
\vec{b}_2 = \begin{pmatrix} 0 \\[1ex] \frac{2 \pi}{a b \sin \theta '} \end{pmatrix}.
\end{align}
The symmetric structure of $\theta ' = \frac{\pi}{3}$ and $b = 1$
shown in Fig.~\ref{fig: trivistruk} then corresponds to the reciprocal
lattice in the upper part of Fig.~\ref{fig: reci trivi}.  Three different
symmetry points are immediately recognized and marked by $\Gamma$, X
and J in the figure.  They are respectively the point
$(0,0)$, the point between the two dipoles at $(0,0)$ and
$\vec{b}_1+\vec{b}_2$, and the corner between three Brillouin zones
for the dipoles at $(0,0)$, $\vec{b}_1$ and $\vec{b}_1+\vec{b}_2$.

Increasing the value of $\theta'$ above $\pi/3$ moves the
$\vec{b}_1$-vector closer to the $x$-axis.  This results in a twisting
of the reciprocal lattice as shown in lower part of Fig. \ref{fig:
  reci trivi}.  The symmetry points, $\Gamma$, X and J, for the
perpendicular dipoles solution are also present for all other values
of $b$ and $\theta'$.  Decreasing $\theta'$ would lead to a twisting
of the reciprocal lattice in the opposite direction of going from
upper to lower part of Fig. \ref{fig: reci trivi}.

These symmetry points will be natural target points for the wave
vector in investigations of the phonon spectra of the different
structures.

\subsection{Results for perpendicular dipoles: $\theta=0$}

The phonon spectra are calculated along the closed path passing
through symmetry points $\Gamma$, $X$ and $J$ shown in upper panel on
Fig. \ref{fig: reci trivi}.  The wave vector, $\vec{k}$, always begins
at the origin, $\Gamma$.  First it increases in size in the direction
of $X$, then it follows the line towards $J$, and finally it maintains
this direction while decreasing size until zero at the starting
point, $\Gamma$.

\begin{figure}
\centering
\includegraphics[scale=0.5]{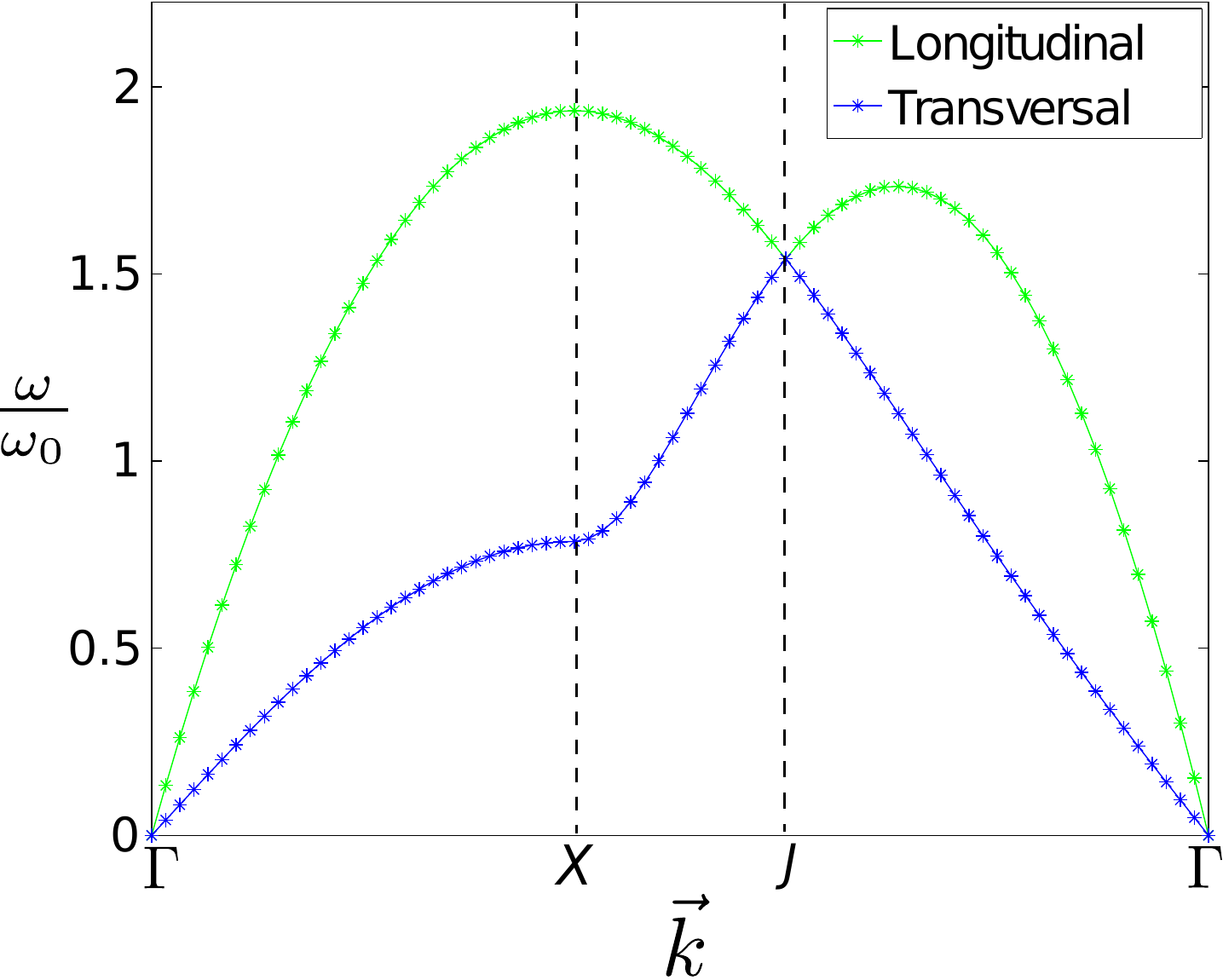}
\caption{The phonon frequency in units of $\omega_0 = \sqrt{\frac{D^2
    \mu_0 \sigma ^{5/2}}{M}}$ for
  perpendicular dipoles of $\theta = 0$ as function of the path
  followed by the wave vector, $k$. The direction of $\vec{k}$
  corresponds to the path with end-points marked in Fig. \ref{fig:
    reci trivi}, that is from $\Gamma$ over $X$, towards $J$ and back
  again to $\Gamma$.  The relative sizes on the $x$-axis are
  arbitrary.  }
\label{fig: rotation}
\end{figure}

The phonon spectrum is then calculated and shown in Fig. \ref{fig:
  rotation} for perpendicular dipoles of $\theta=0$.  Two frequencies
arise from diagonalizing the $2 \times 2$ $D$-matrix. The two types of
vibrations are denoted longitudinal and transverse corresponding to
small amplitude motion of all dipoles along and perpendicular to the
$\vec{k}$-direction, respectively.

The two computed frequencies both increase from zero at the symmetry
point, $\Gamma$.  The longitudinal mode show the fastest increase of
energy, reaching a maximum at the edge of the Brillouin zone, when the
wave vector reaches its maximum at the symmetry point $X$.  The
transverse frequency increases slower but with a similar flat region
before reaching $X$.  

With the wave vector end-point continuing from $X$ towards $J$ we find
a smoothly decreasing longitudinal frequency and a more abruptly, yet
continuously, increasing transversal frequency.  

As the symmetry point, $J$, for the hexagonal lattice is approached
the two frequencies meet each other in one degenerate value. 
As this is a highly symmetrical point for the hexagonal lattice such
a behavior would be expected here but it is also only expected for this particular case.

The two frequency curves continue smoothly through this point, but
interchange correspondingly vibrational character between transversal
and longitudinal mode.  
The longitudinal mode continues to reach a
maximum before decreasing towards zero at $\Gamma$, while the
transversal mode decreases directly towards zero at the central point.

Both frequencies approach zero at the symmetry point, $\Gamma$,
although with different rates.  Thus no optical branch is found in
agreement with what is expected for a single particle basis.

\subsection{Results for tilted dipoles: $0.1 \leq \theta \leq 0.6  $}

\begin{figure}
\centering
\includegraphics[scale=0.5]{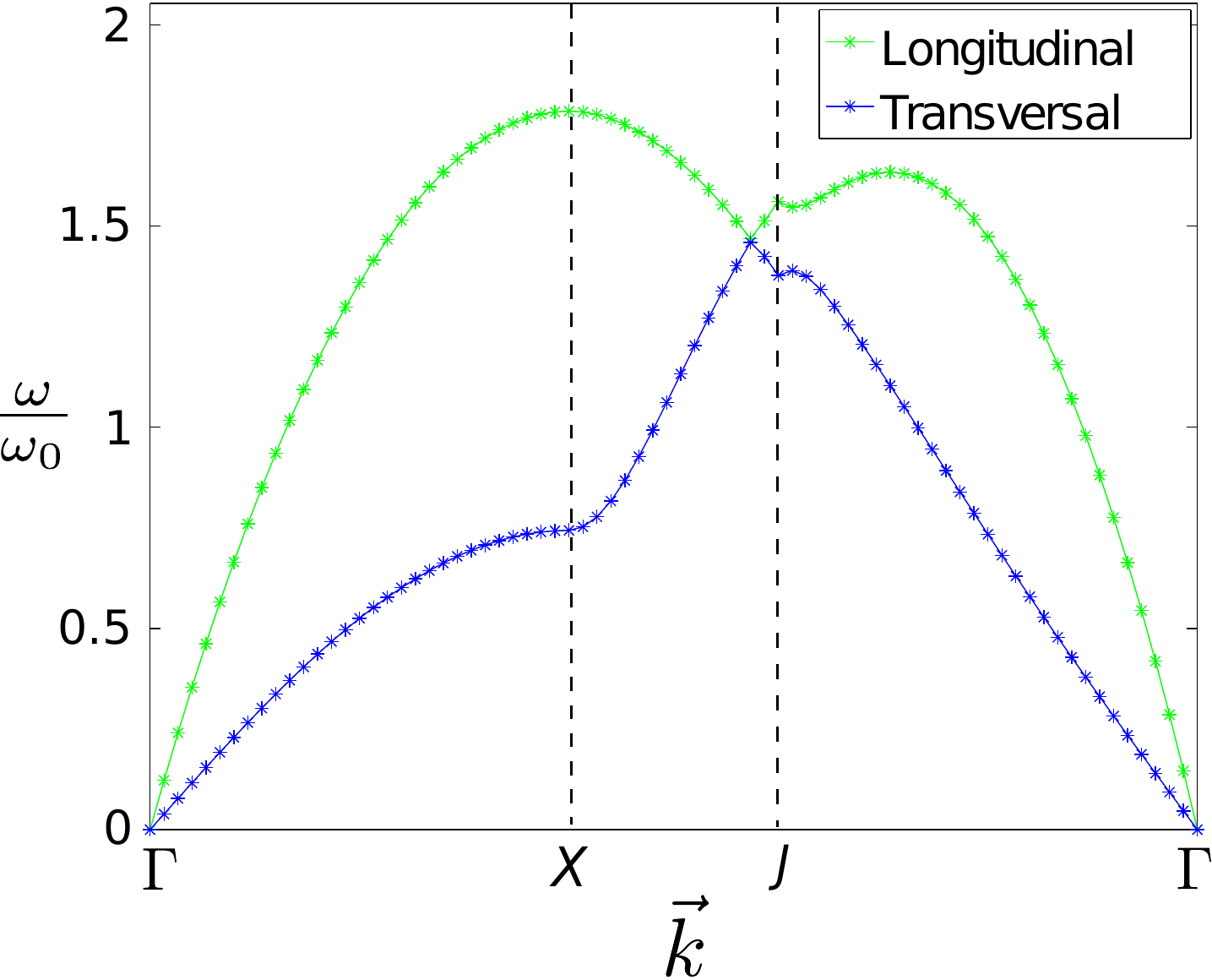} \\
\includegraphics[scale=0.5]{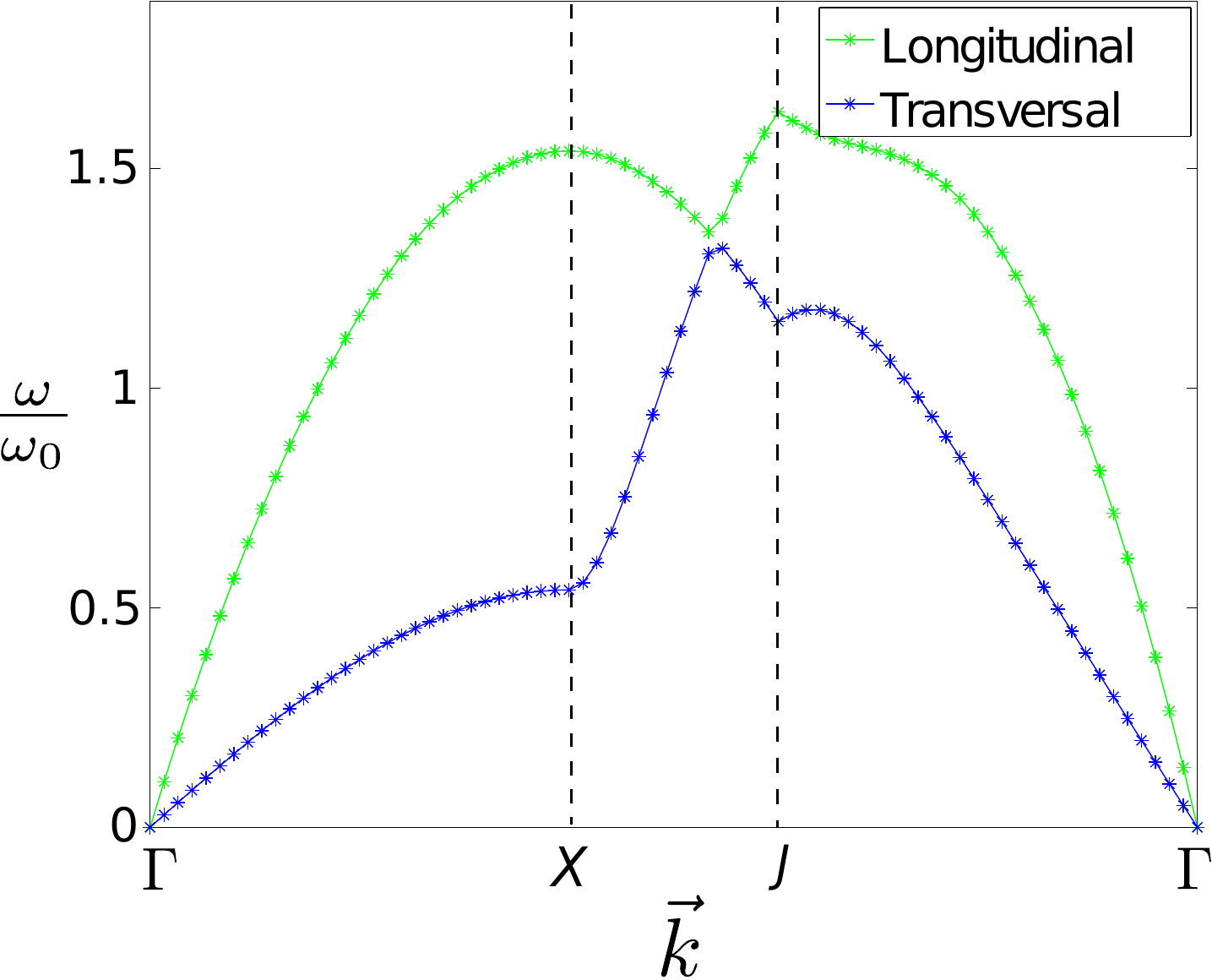} \\
\includegraphics[scale=0.5]{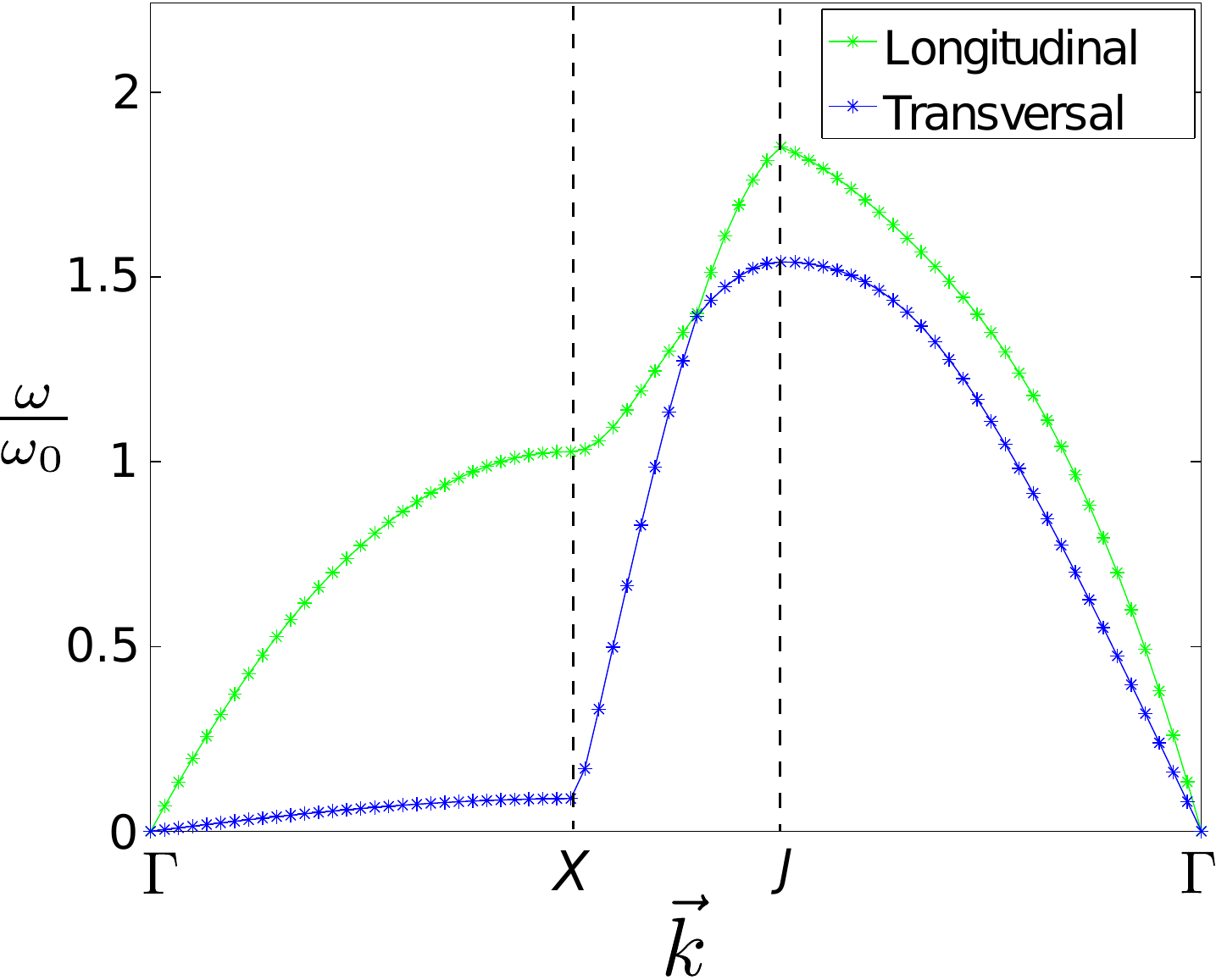}
\caption{The phonon spectra for tilted dipoles with polarization
  directions corresponding to $\theta=0.25$ (upper panel), $\theta=0.40$ (middle panel), and $\theta=0.55$ (lower panel).}
\label{fig: spektre-0,1}
\end{figure}

We now slowly increase $\theta$ above zero.  For small values the
structure is rather similar to that of $\theta = 0$.  Accordingly the
changes in the phonon spectrum is small although noticeable.  The
complete hexagonal symmetry is broken and a $\phi$-dependence of the
interaction energy appear for finite values of $\theta$, see
Fig. \ref{fig: cirkel0,1}.  The symmetry is no longer perfect even for
a small value of $\theta = 0.1$, but the symmetry points remains in
the reciprocal lattice sketched in Fig. \ref{fig: reci trivi}.  The
degeneracy in the point $J$ is lifted and the two branches begin to
move away from each other at this point.

Increasing $\theta$ changes the structure substantially especially
towards the critical value $\theta_c$.  The results for increasing
$\theta$ are shown in Fig. \ref{fig: spektre-0,1} for three larger
illustrative values.  For $\theta = 0.25$ the degeneracy at point $J$
has now clearly disappeared and a significant gap has appeared.  The
spectrum still maintain the same features and the same overall
appearance.  However, now a small minimum appears on the longitudinal
branch close to the point $J$.  This minimum is present for all
spectra in the interval from $\theta = 0.2-0.35$.  After appearance it
first becomes deeper but as $\theta$ grows so does the frequency at
$J$ and for $\theta = 0.35$ the minimum is altogether washed out.  The
significance of this minimum could possible be a signal of roton
dynamics in the crystal in analogy to that of helium \cite{helium}.
Such rotons have been predicted and discussed also for particles 
with dipolar interactions 
\cite{mazzanti2009,hufnagl2011,zinner2012,macia2012,bisset2013,fedorov2014a,fedorov2014b}.
It is clearly related to deformation of the crystal, but we postpone
more detailed investigations to future work.

Increasing $\theta$ to $0.4$ and beyond, we see that the spectra in
Fig. \ref{fig: spektre-0,1} change appearance.  The point $X$ moves
closer to $\Gamma$ in Fig. \ref{fig: reci trivi}.  We note that no new
minima appear beside the kink at $J$ in the transversal frequency for
$\theta=0.4$.  The spectrum for $\theta = 0.55$ is apparently more
distorted with kinks or abruptly changing values.  However, we
emphasize that the $x$-axes on Fig. \ref{fig: spektre-0,1} do not
reflect real distances.  It is merely the result of the choice of
discrete wave vectors $\vec{k}$, and the subsequent distance between
points on the figures.  Therefore the variation is not a realistic
signal representative for any physical effect.

\begin{figure}
\centering
\includegraphics[scale=0.5]{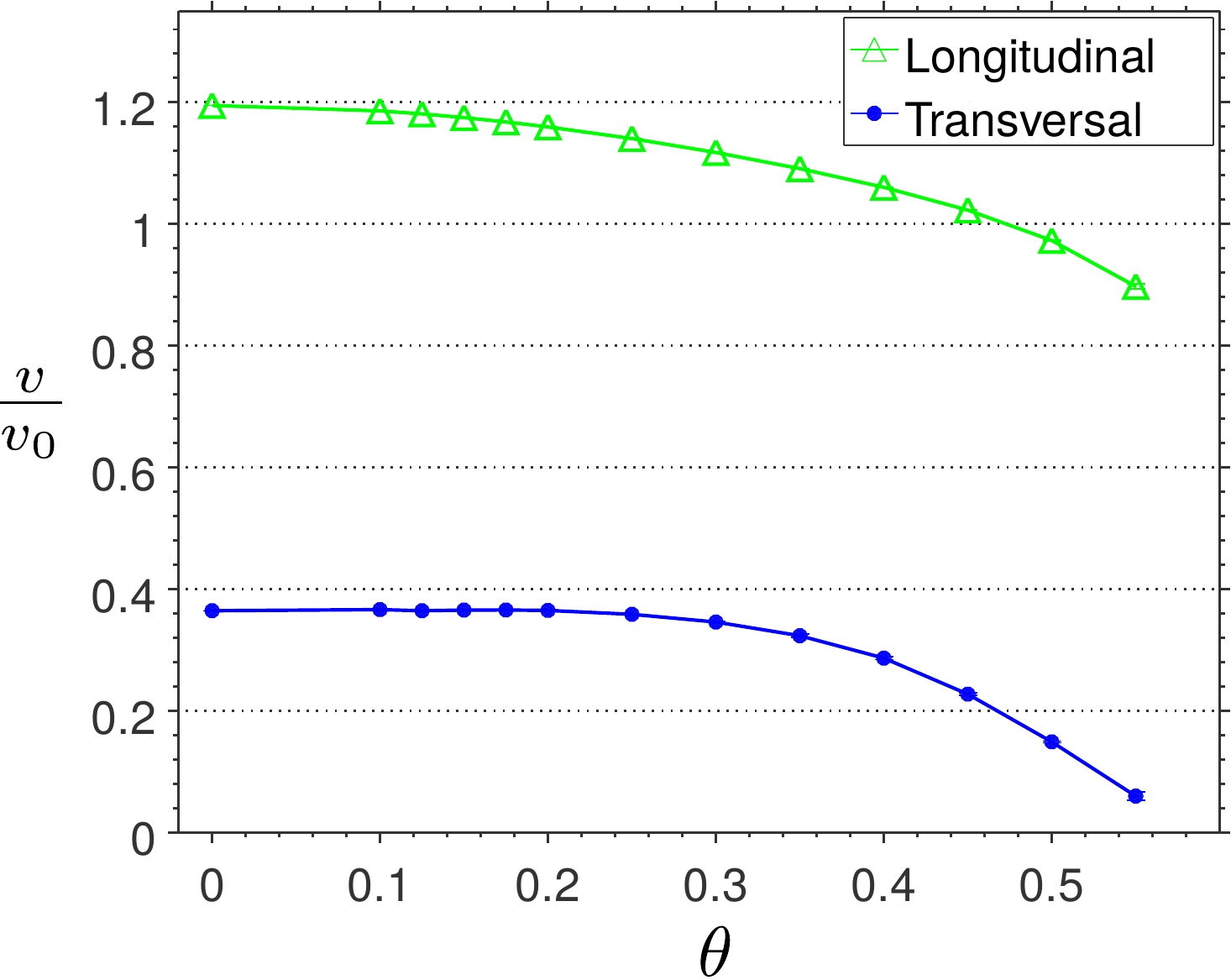}
\caption{The speed of sound as a function of $\theta$. Here $v_0 = \sqrt{\frac{D^2 \mu_0 \sigma^{3/2}}{M}}$.}
\label{fig: sound}
\end{figure}

\subsection{Sound velocities}
With the phonon spectrum available we calculate the speed of sound,
$v$, from Eq.~\eqref{eq:lyd} in the point, $\Gamma$, in both transverse
and longitudinal directions.  The results are seen to be the
derivatives of the frequency curves in Figs. \ref{fig: rotation} and
\ref{fig: spektre-0,1}.  Each configuration then has longitudinal and
transversal values for the speed of sound as shown in in
Fig.~\ref{fig: sound} as functions of $\theta$.  For the hexagonal
structure, $\theta=0$, we find $v = 1.19481 v_0$ and $v = 0.364272
v_0$, respectively, in the natural units of 
$v_0 = \sqrt{\frac{D^2\mu_0 \sigma ^{3/2}}{M}}$.  

Both longitudinal and transversal velocities decrease continuously and
relatively slowly with $\theta$.  There is a tendency to speed up the
reduction as the critical polarization angle is approached.  This is
intuitively understandable, since the structure for $\theta
\rightarrow \theta_c$ approaches separated straight lines as seen in
Fig. \ref{fig: struktur-0-0,6}.  The frequencies decrease when the
instability is approached, and correspondingly the speed of sound must
decrease. This instability in also clear to see in bottom panel 
of Fig.~\ref{fig: spektre-0,1} where we see the normal mode going 
towards zero energy
at the point $X$.

\section{Summary and Outlook}\label{sum}

Using a classical analysis we have determined the structure, the
phonon spectrum, and the speed of sound for dipolar crystals in two
dimensions with different polarization of the dipole moments with
respect to the two-dimensional plane of motion. This analysis has been
restricted to polarization angles less than the critical angle for
dipole-dipole collapse.  This ensures that the overall interaction is
repulsive so that one avoids collapse due to head-to-tail attraction
of the dipoles.

For perpendicular orientation, we find the expected hexagonal crystal
structure. As the polarization angle increases, we observe a
distortion of the crystal lattice and the structure changes towards a
more square type of lattice.  This is associated with the increasing
tendency for the system to prefer stripes of dipoles.  The physics of
this deformation is associated with the decreasing repulsive energy
felt by two dipoles as they are tilted away from perpendicular
orientation. The optimal energy configuration therefore becomes one
where dipoles are placed on lines and effectively makes a striped
system. Indications of stripe formation have also been found in
different response function approaches \cite{sun2010,zinner2011,parish2012}.

As we have demonstrated above, the transition from hexagonal toward a
striped character of the system is particularly clear in the phonon
spectrum and in the speed of sound of the system. Measurements of
phonon dispersions and speeds of sound would therefore be a promising
way to experimentally probe the crystal.  Interestingly, we find that
the phonon spectrum for increasing tilt angles develops some local
minima analogous to the roton minima seen in Helium.  These new minima
occur in positions different from the ones found in the case of
perpendicular orientation and are thus associated with the deformation
of the crystal structure.

Our study provides the basis for including quantum mechanical effects
in the system. This becomes important once the dipolar interaction
strength is comparable to the kinetic energy of the zero-point motion
in the crystal which occurs for weak dipole moments. When the
zero-point energy is large we expect strong quantum fluctuations and
suppression of crystal formation. Including these effects, one can
investigate melting transitions and heat capacity due to quantum
motion.  This can be done by using canonical quantization of the
phonon modes and/or via a Lindemann criterion approach.  This will be
the subject of future work.

\appendix
\section{Derivation of the Phonon Spectrum}\label{appa}
The derivation will follow the derivation in chapter 22 in Ashcroft and Mermin (1976)\citep{faststof2}.

As mentioned before the energy of a structure is calculated as
\begin{equation}
I = \frac{1}{2} \sum_{\substack{
\vec{R}\vec{R'} \\ \vec{R} \neq \vec{R'}}}
\phi (\vec{R}-\vec{R'}).
\end{equation}
Here $\vec{R}$ and $\vec{R'}$ are the position vectors of each dipole in the structure and $\phi$ is the interaction potential between two dipoles.
We introduce a small perturbation in the system as
\begin{equation}
\vec{r}(\vec{R}) = \vec{R} + \vec{u}(\vec{R}).
\end{equation}
Insert this in the expression for the energy,
\begin{align}
I &= \frac{1}{2} \sum_{\substack{
\vec{R}\vec{R'} \\ \vec{R} \neq \vec{R'}}}
\phi (\vec{r}(\vec{R})-\vec{r}(\vec{R'}))\\
&= \frac{1}{2} \sum_{\substack{
\vec{R}\vec{R'} \\ \vec{R} \neq \vec{R'}}}
\phi (\vec{R} - \vec{R'} + \vec{u}(\vec{R}) - \vec{u}(\vec{R'})).
\end{align}
An expansion of the potential is made in $\vec{u}(\vec{R})-\vec{u}(\vec{R'})$,
\begin{equation}
f(\vec{r}+\vec{a}) = f(\vec{r}) + \vec{a}\cdot\nabla f(\vec{r})+ \frac{1}{2!}(\vec{a}\cdot\nabla)^2 f (\vec{r}) \cdots
\end{equation}
All terms of order $O(\vec{u}^3)$ are removed as a harmonic approximation a is made. This is appropriate as vibrations should be small for zero Kelvin.
The zeroth order term will in principle be an infinite term but it will not be relevant for motion. The linear term will not be relevant as we assume all particles to be at minimum positions and will therefore be zero.
The Harmonic term,
\begin{align}
\begin{split}
I^{\mathrm{Harm}} = \frac{1}{4} \sum_{
\substack{\vec{R}\vec{R'} \\ \mu,\nu = x,y \\ \vec{R} \neq \vec{R'}}}
&\times[u_\mu (\vec{R}) - u_\mu(\vec{R'})]\\&\frac{\partial ^2 \phi (\vec{R}-\vec{R'})}{\partial r_\mu \partial r_\nu} [u_\nu (\vec{R}) - u_\nu (\vec{R'})].
\end{split}
\end{align}
can be rewritten to something more useful
\begin{equation}
\begin{split}
I^{\mathrm{Harm}} = \frac{1}{4} \sum_{\substack{\vec{R}\vec{R'}\\ \mu,\nu = x,y \\ \vec{R}\neq\vec{R'}}}
&(u_{\mu} (\vec{R}) \phi_{\mu \nu} (\vec{R}-\vec{R'}) u_\nu (\vec{R})\\
&+u_{\mu} (\vec{R'}) \phi_{\mu \nu} (\vec{R} - \vec{R'}) u_{\nu} (\vec{R'})\\
&-u_{\mu} (\vec{R}) \phi_{\mu \nu} (\vec{R}-\vec{R'}) u_{\nu} (\vec{R'})\\
&-u_{\mu} (\vec{R'}) \phi_{\mu \nu} (\vec{R}-\vec{R'}) u_{\nu} (\vec{R})),
\end{split}
\end{equation}
where $\phi_{\mu \nu} (\vec{R}-\vec{R'}) = \frac{\partial ^2 \phi (\vec{R}-\vec{R'})}{\partial r_\mu \partial r_\nu}(\vec{R}-\vec{R'})$. 
Because $\phi_{\mu \nu} (\vec{R}-\vec{R'}) = \phi_{\mu \nu} (\vec{R'}-\vec{R}) \Rightarrow u_{\mu} (\vec{R}) \phi_{\mu \nu} (\vec{R}-\vec{R'}) u_\nu (\vec{R}) = u_{\mu} (\vec{R'}) \phi_{\mu \nu} (\vec{R} - \vec{R'}) u_{\nu} (\vec{R'})$, and because $\vec{R}$ and $\vec{R'}$ express the same points it will, per symmetry, be that $u_{\mu} (\vec{R}) \phi_{\mu \nu} (\vec{R}-\vec{R'}) u_{\nu} (\vec{R'}) =
u_{\mu} (\vec{R'}) \phi_{\mu \nu} (\vec{R}-\vec{R'}) u_{\nu} (\vec{R})$.
The harmonic term will again be rewritten as
\begin{equation}
\begin{split}
I^{\mathrm{Harm}} = \frac{1}{2} \sum_{\substack{\vec{R}\vec{R'}\\ \mu,\nu = x,y \\ \vec{R}\neq\vec{R'}}}
&u_{\mu} (\vec{R}) \phi_{\mu \nu} (\vec{R}-\vec{R'}) u_\nu (\vec{R})\\
&-u_{\mu} (\vec{R}) \phi_{\mu \nu} (\vec{R}-\vec{R'}) u_{\nu} (\vec{R'}).
\end{split}
\end{equation}
A new notation is introduced so only one term is present:
\begin{equation}
I^{\mathrm{Harm}} = \frac{1}{2} \sum_{\substack{\vec{R}\vec{R'}\\ \mu,\nu = x,y}}
u_{\mu} (\vec{R}) D_{\mu \nu} (\vec{R}-\vec{R'}) u_\nu (\vec{R'}),
\label{eq: D(R)} 
\end{equation}
where $D_{\nu \mu} (\vec{R}-\vec{R'}) = \delta_{\vec{R},\vec{R'}} \sum_{\vec{R''}} \phi_{\nu \mu} (\vec{R}-\vec{R''}) - \phi_{\mu \nu} (\vec{R} - \vec{R'})$ and the sum allows $\vec{R}=\vec{R'}$.
Introduce a matrix notation instead:
\begin{equation}
I^{\mathrm{Harm}} = \frac{1}{2} \sum_{\vec{R}\vec{R'}}
\vec{u} (\vec{R}) \underline{\underline{D}}(\vec{R}-\vec{R'}) \vec{u}(\vec{R'}).
\end{equation}
This matrix is symmetric as a result of $\frac{\partial ^2 \phi (\vec{R}-\vec{R'})}{\partial r_x \partial r_y} = \frac{\partial ^2 \phi (\vec{R}-\vec{R'})}{\partial r_y \partial r_x}$.

It it also clear that $D_{\mu \nu} =(\vec{R}-\vec{R'})$ can be expressed as:
\begin{equation}
D_{\mu \nu} (\vec{R}-\vec{R'}) = \left. \frac{\partial^2 I}{\partial u_{\mu} (\vec{R}) \partial u_\nu (\vec{R'})}\right|_{u \equiv 0}.
\end{equation}
Now two equations of motion can be made for each particle
\begin{equation}
M \ddot{u}_\mu (\vec{R}) = F_\mu = -\frac{\partial I^{\mathrm{Harm}}}{\partial u_\mu (\vec{R})} = -\sum_{\vec{R'} \nu} D_{\mu \nu} (\vec{R} - \vec{R'}) u_\nu (\vec{R'}).
\end{equation}
Here the differential has removed two sums in the expression, M is the mass of each dipole and $F_\mu$ is the force in the $\mu$'th direction.
These equations can then be collected in vector notation as
\begin{equation}
M \vec{\ddot{u}} (\vec{R}) = - \sum_{\vec{R'}} \underline{\underline{D}} (\vec{R}- \vec{R'}) \vec{u}(\vec{R'}).
\end{equation}
Solving these equations require us to make the ansatz:
\begin{equation}
\vec{u}(\vec{R},t) = \vec{\epsilon} e^{i(\vec{k} \cdot \vec{R} -\omega t)},
\end{equation}
where $\vec{k}$ is a wavevector.
This results in:
\begin{align}
M \ddot{u}_{\mu} (\vec{R}) = & -M \omega ^2 \vec{\epsilon} e^{i(\vec{k} \cdot \vec{R} -\omega t)}\\
= & -\sum_{\vec{R'}} \underline{\underline{D}} (\vec{R}- \vec{R'}) \vec{u} (\vec{R'},t)\\
= & -\sum_{\vec{R'}} \underline{\underline{D}} (\vec{R}- \vec{R'}) \vec{\epsilon} e^{i(\vec{k} \cdot \vec{R} -\omega t)}\\
= & -\vec{\epsilon} e^{-i \omega t} \sum_{\vec{R'}} \underline{\underline{D}} (\vec{R}- \vec{R'}) e^{i \vec{k} \cdot \vec{R'}} \\
\Rightarrow \\
M \omega ^2 \vec{\epsilon} =& \vec{\epsilon} \sum_{\vec{R'}} \underline{\underline{D}} (\vec{R} - \vec{R'}) e^{i \vec{k} \cdot (\vec{R'}-\vec{R})} \\
= &\vec{\epsilon} \sum_{\vec{R'}} \underline{\underline{D}} (\vec{R'}- \vec{R}) e^{i\vec{k} \cdot (\vec{R'}-\vec{R})}.
\end{align}
Because both $\vec{R}$ and $\vec{R'}$ describe all positions in the infinite structure we now make the substitution $\vec{R''} = \vec{R} - \vec{R'}$ and get:
 \begin{equation}
M \omega ^2 \vec{\epsilon} =\vec{\epsilon} \sum_{\vec{R''}} \underline{\underline{D}} (\vec{R''}) e^{-i\vec{k} \cdot \vec{R''}} = \underline{\underline{D}} (\vec{k}) \vec{\epsilon},
\end{equation}
where $\underline{\underline{D}} (\vec{k}) = \sum_{\vec{R''}} \underline{\underline{D}} (\vec{R''}) e^{-i\vec{k} \cdot \vec{R''}} = -2 \sum_{\vec{R''}} \underline{\underline{D}} (\vec{R''}) \sin^2(\frac{1}{2} \vec{k}\cdot \vec{R''})$.
The expression has now been reduced to a eigenvalue equation which can be solved for a given $\vec{k}$.
If the structure would be of finite proportions there would be restrictions on the allowed values of $\vec{k}$ but if infinite it could have any value in the first Brillouin zone.
As the matrix has real values, is symmetric and of dimension 2 there will be two real eigenvalues with two orthogonal eigenvectors which fulfils:
\begin{equation}
\underline{\underline{D}} (\vec{k}) \vec{\epsilon_i} (\vec{k}) = \lambda_i (\vec{k}) \vec{\epsilon_i} (\vec{k}).
\label{eq: egen}
\end{equation}
From this $\omega_i (\vec{k})$ belonging to $\vec{\epsilon_i} (\vec{k})$ can be found as
\begin{equation}
\omega_i (\vec{k}) = \sqrt{\frac{\lambda_i (\vec{k})}{M}}.
\end{equation}

\end{document}